\documentclass[lettersize,journal]{IEEEtran}
\usepackage{comment}
\usepackage{amsmath,amsfonts}
\usepackage{amssymb}
\usepackage{cuted}
\usepackage{flushend}
\usepackage{xfrac}
\usepackage{algorithmic}
\usepackage{algorithm}
\usepackage{array}
\usepackage{textcomp}
\usepackage{stfloats}
\usepackage{url}
\usepackage{verbatim}
\usepackage{graphicx}
\usepackage{cite}
\usepackage{mathrsfs}
\usepackage{mathtools,lipsum}

\usepackage{hyperref}
\usepackage[abs]{overpic}
\usepackage{amsthm}
\newtheorem{thm}{Theorem}
\newtheorem{lem}{Lemma}

\theoremstyle{definition} 
\newtheorem{rem}{Remark}

\usepackage{subcaption}
\usepackage{cuted}   
\usepackage{titlesec}

\usepackage[font=small,skip=0pt]{caption}

\usepackage{multicol}

\begin{document}
\title{Massive MIMO over Correlated Fading Channels: Multi-Cell MMSE Processing, Pilot Assignment and Power Control}
\author{Masoud Elyasi, Azadeh Vosoughi,~\IEEEmembership{Senior Member,~IEEE}\thanks{Author Contributions: The novelty of the work, formula derivation, and simulations were developed by Masoud Elyasi (M4soud.elyasi@gmail.com). Azadeh Vosoughi contributed by reviewing and editing the manuscript for clarity and technical accuracy. This work was conducted voluntarily by the authors without institutional support or external funding.}}

\maketitle
\begin{abstract}
We consider a multi-cell massive multiple-input multiple-output (MIMO) system operating under spatially correlated Rayleigh fading channels, where pilot reuse is permitted both within and across cells, and each base station (BS) employs multi-cell minimum mean square error (M-MMSE) processing. We derive a novel deterministic approximation of the uplink signal-to-interference-and-noise ratio (SINR), asymptotically tight in the large-system limit, even under pilot reuse and spatial correlation, addressing a key gap in the existing literature. Building on this result, we propose a multi-cell pilot assignment (PA) scheme that fully eliminates pilot contamination by exploiting the spatial correlation matrices of all users. To ensure scalability in large networks, we further introduce a scalable PA scheme with partial M-MMSE (P-MMSE) processing, which reduces inter-BS information exchange while maintaining high spectral efficiency (SE). Additionally, we design pilot and data power allocation strategies for both weighted sum SE and max-min SE objectives. A detailed complexity analysis confirms the practicality of the proposed algorithms. Simulation results demonstrate the robustness and superiority of our PA schemes across various network conditions, showing substantial SE gains and good user fairness with significantly lower pilot overhead compared to existing approaches, offering valuable insights for the design of future massive MIMO systems.\end{abstract}
\section{Introduction}\label{sec: Introduction}
\IEEEPARstart{M}{assive} Multiple-Input Multiple-Output (MIMO) technology has become a cornerstone of modern wireless communication systems, playing a critical role not only in 5G but also in the evolution toward 6G networks. Its integration with emerging paradigms such as intelligent reflecting surfaces (IRS) \cite{wang2022massive,papazafeiropoulos2023achievable,nguyen2022hybrid}, wireless information and power transfer (WIPT) \cite{amudala2023wireless}, and integrated sensing and communication (ISAC) \cite{liao2024power,zhou2024integrating} further highlights its importance. The core principle of massive MIMO involves equipping cellular base stations (BSs) with a large number of antennas to simultaneously serve multiple users over the same time-frequency resources. This enables significant spectral efficiency (SE) gains by averaging out intra-cell interference and uncorrelated noise through the use of linear precoders and detectors \cite{andrews2014will}.

Most existing literature on massive MIMO focuses on single-cell processing architectures, where each BS detects and precodes only for its own intra-cell users using methods such as matched filtering (MF), zero-forcing (ZF), or single-cell MMSE (S-MMSE) processing \cite{marzetta2016fundamentals}\cite{hoydis2013massive}. These works often assume uncorrelated Rayleigh fading models and no pilot reuse within a cell, which simplifies the derivation of closed-form SE expressions and power allocation schemes \cite{wang2022massive,xie2023performance,wang2023uplink,dey2023rsma}. However, these simplifications come at the cost of limiting realism and performance. In contrast, multi-cell processing schemes, especially the multi-cell MMSE (M-MMSE) detector/precoder introduced in \cite{jose2011pilot,bjornson2017massive}, exploit channel estimates from all users across the network and are shown to achieve superior SE by mitigating inter-cell interference.

Despite these advances, key gaps remain in the literature. Notably, the closed-form SE expressions derived for M-MMSE processing in prior works \cite{li2017massive} rely on the uncorrelated Rayleigh fading assumption and do not permit pilot reuse within cells. These assumptions restrict the applicability of such methods to realistic massive MIMO deployments, where spatial correlation among antennas and pilot reuse due to limited pilot resources are prevalent. Furthermore, pilot contamination caused by pilot reuse remains the fundamental bottleneck in multi-cell massive MIMO systems, as increasing the number of antennas alone cannot eliminate this effect \cite{marzetta2010noncooperative}, \cite{marzetta2016fundamentals}.

Various pilot assignment (PA) strategies have been proposed to mitigate pilot contamination in massive MIMO systems. For example, the smart PA scheme in \cite{zhu2015smart} enhances max-min fairness SE by optimizing a heuristic metric based on large-scale fading characteristics between pilot-sharing users in adjacent cells. Building on this idea, the weighted graph-based PA scheme in \cite{zhu2016weighted} improves uplink sum SE by quantifying the strength of potential pilot contamination between users and allocating pilots accordingly. This weighted graph approach has been further extended in several recent works \cite{liu2020graph, xia2024graph, kumar2024scalable, hu2024graph}, particularly within the cell-free massive MIMO framework, where a large number of distributed access points jointly serve users without cell boundaries. Another notable line of work is \cite{yin2013coordinated}, which considers spatial correlation matrices and shows that if pilot-sharing users exhibit non-overlapping angle-of-arrival (AoA) intervals, channel estimation can be made free of pilot contamination. However, the scheme in \cite{yin2013coordinated} focuses on minimizing the sum of channel estimation error covariances and requires global access to all user covariance matrices across the network—introducing significant overhead. Moreover, its assumption of small AoA spreads may not hold in many practical deployment scenarios.

It is important to emphasize that all of these existing PA schemes were developed under the assumption of single-cell processing architectures, where each BS employs precoders and detectors based solely on its own intra-cell channel estimates. These single-cell PA designs do not account for the joint impact of pilot contamination across multiple cells when M-MMSE processing is applied, which leverages channel estimates from all users in the network. The approach in \cite{li2017massive} considers M-MMSE processing under uncorrelated Rayleigh fading and mitigates pilot contamination by introducing a pilot reuse factor $f$, where each pilot is reused in only a fraction of the cells. While this reduces pilot contamination, it comes at the cost of increasing pilot overhead by a factor of $f$, requiring large coherence block lengths (e.g., $\tau_c = 1000$ or $2000$ symbols) to remain effective. However, under such large coherence block conditions, blind channel estimation methods \cite{hu2015semi,muller2014blind,rekik2024fast} could instead be employed to suppress pilot contamination without additional pilot overhead, limiting the practicality of reuse-based schemes like \cite{li2017massive} in dynamic outdoor environments.

In this work, we consider a multi-cell massive MIMO network operating under spatially correlated Rayleigh fading, where pilot reuse is permitted both within and across cells, and where M-MMSE processing is employed at the BSs. We identify and address the key theoretical and practical challenges that arise in this setting, particularly those related to pilot contamination mitigation, channel estimation accuracy, and scalability under limited pilot resources. The main contributions of this paper are summarized as follows:

\noindent
$\bullet$ By applying tools from random matrix theory, we derive a new deterministic approximation of the uplink SINR for M-MMSE processing under spatially correlated Rayleigh fading (Theorem 3). This result fills an important gap in the literature, where closed-form SINR approximations have been unavailable under such conditions, even in cell-free massive MIMO systems. The derivation involves significant analytical challenges due to the combined effects of M-MMSE processing, spatial correlation, and pilot reuse. As discussed in Remark 3 (following Theorem 3), these challenges make the proof substantially more complex than the corresponding derivations in \cite{li2017massive}, which consider uncorrelated fading. To the best of our knowledge, this SINR approximation and the associated SE expression represent the first results of this kind for correlated massive MIMO systems.


\noindent
$\bullet$ Building on the derived SINR approximation, we propose a novel multi-cell PA algorithm (Algorithm 1) that fully eliminates pilot contamination by minimizing coherent interference between pilot-sharing users, leveraging the spatial correlation matrices of all users across the network. The pilot assignment is guided by a new optimality condition (Theorem 4). To ensure scalability in large networks, we also introduce Algorithm 2, a scalable extension of the multi-cell PA scheme based on partial M-MMSE (P-MMSE) processing. This scalable scheme limits inter-BS information exchange by considering only a subset of users, selected via a tunable threshold on their channel strength, while maintaining near-optimal SE as the network size grows. We further provide a detailed computational complexity analysis, and show that the information exchange overhead of the scalable scheme remains bounded and independent of the number of antennas, making it well-suited for large-scale deployments.


\noindent
$\bullet$  Building on the proposed PA framework, we formulate and solve a pilot power optimization problem to minimize the weighted sum of channel estimation error covariances across the network. Additionally, we derive the optimal uplink data power allocation strategy using the deterministic SINR approximation, and by leveraging uplink-downlink duality, we determine the optimal downlink data power allocation to maximize both sum SE and max-min SE objectives.

\noindent
$\bullet$ Through extensive simulations, we demonstrate the superiority of the proposed multi-cell PA schemes over existing PA  methods across a range of scenarios, including different detector types, angular spreads, numbers of cells, antenna counts, pilot lengths, and spatial correlation conditions. The results confirm the robustness, scalability, and practical applicability of our approach, offering valuable system design insights for real-world deployments.


\textit{Notation}: Upper and lower case bold letters are used for matrices and vectors. The superscripts ${()}^T$, ${()}^H$ stand for transpose and Hermitian operations, respectively. The $\mathbb{R}$ and $\mathbb{C}$ are adopted for real value and complex numbers. The expectation, variance, trace and diagonal of a matrix, real part of the argument, Euclidean norm and Frobenius norm are denoted by $\mathbb{E}\left\{\ \right\}$, $\mathbb{V}\left\{\ \right\}$, $tr(.)$, $diag(.)$, $\Re\left\{\ \right\}$, $\| \|_2$ and $\| \|_F$, respectively. $\mathbf{I}_M$ and $\mathbf{0}_M$ is the size-$M$ identity and all-zero matrix, respectively. We use $\mathcal{N}_\mathbb{C}\left(\mathbf{0},\mathbf{R}\right)$ to denote the circularly symmetric complex Gaussian distribution with zero mean and covariance matrix {\bf{R}}. Also we use $a\asymp b$ to denote $a-b\stackrel{{\text{a.s.}}}{\longrightarrow}0$ almost surely (a.s.) for two random sequences. User $jk$ refers to user $k$ in cell $j$. 

The rest of the paper is organized as follows. In Section \ref{sec: System Model}, we describe the system model, derive achievable uplink and downlink rates, and provide an intuition for the M-MMSE scheme. In Section \ref{sec: Achievable Rates}, we derive an asymptotically tight approximation of the uplink SINR. In Section \ref{sec:pilot assignment}, we characterize  our proposed multi-cell and single-cell PA schemes. 
Section \ref{sec: Power Control} details our proposed pilot and data power allocation. Section \ref{sec: Simulation results and discussion} includes our numerical results. Section \ref{CONCLUSION} contains our concluding remarks. 

\section{System Model and Transceiver Design}\label{sec: System Model}
We consider a multi-cell massive MIMO system with $L$ cells indexed by $l=1,\dots, L$. Each cell has a base station (BS) with $M$ antennas and serves $K (K\ll\ M)$ single-antenna users, indexed by $k=1,...,K$, over a time-frequency coherence block. We adopt the standard block-fading channel model, where the channel stays constant over each coherence block, and the channels across blocks are independent and identically distributed (i.i.d.). Each coherence block corresponds to $B_c$ Hz and $T_c$ seconds, where $B_c$ is smaller than the coherence bandwidth of all users and $T_c$ is smaller than the coherence time of all users, and contains $\tau_c=B_c\times T_c$ transmission symbols. The channel vector $\mathbf{h}_{lk}^j \epsilon{\ \mathbb{C}}^{M\times1}$ from user \textit{lk} to BS \textit{j} within a coherence block is modeled as
\begin{equation}
	\label{system_model_eq1}
	\mathbf{h}_{lk}^j \sim \mathcal{N_C}(\mathbf{0},\,\mathbf{R}_{lk}^j)
\end{equation}
where spatial correlation matrix $\mathbf{R}_{lk}^j \in\ \mathbb{C}^{M\times M}$, known at BS $j$ for all $l$ and $k$, depends on the large scale fading coefficient $\beta_{lk}^j$ and the antenna array structure\footnote{In Section IV we describe how ${\bf{R}}_{lk}^j$ is characterized in terms of $\beta_{lk}^j$ and the antenna array structure.}. We assume the system operates based on the TDD protocol, and we exploit channel reciprocity to estimate the downlink channels at each BS using the received uplink pilot signals and employ these channel estimates to process both the received uplink and the transmitted downlink data signals. We assume transmission in each coherence block consists of three phases: 1) uplink training phase, where each user transmits a pilot sequence consisting of $\tau_p$ pilot symbols, where $\tau_p \leq LK$, and each BS estimates the channel vectors of all users of the network, and 2) uplink data transmission phase, where each user transmits $\tau_u$ data symbols, also each BS uses the channel estimates to compute detector and precoder, and 3) downlink data transmission phase, where each BS transmits $\tau_d$ data symbols, using the previously computed precoders. Clearly, we have $\tau_p+\tau_u+\tau_d=\tau_c$.

\subsection{Uplink Channel Estimation}
We assume that $\tau_p$ orthogonal pilot sequences ${\boldsymbol\varphi}_i$, $i=1,...,\tau_p$, of length $\tau_p$ are shared among all $LK$ users of the network, i.e., pilot reuse (sharing) within a cell is allowed. Let ${\boldsymbol\Phi}=[{\boldsymbol\varphi_1},\ldots,{\boldsymbol\varphi_{\tau_p }}]\in\mathbb{C}^{\tau_p\times\tau_p}$, where $||{\boldsymbol\varphi_i}||^2=\tau_p$ and $<{\boldsymbol\varphi_i},{\boldsymbol\varphi_j}>=0$ for $i \neq j$, denote the network pilot matrix. A multi-cell PA scheme (to be discussed in Section IV) assigns a pilot sequence to each user of the network. Suppose $t_{jk} \in \{1,2,...,\tau_p\}$ denote the index of the pilot sequence assigned to user $jk$. Let $\boldsymbol{\varphi}_{t_{jk}}$ represent the pilot sequence transmitted by user $jk$ and ${\hat{p}}_{jk}$ be the corresponding pilot power. Let $\pi_{jk}$ be the set of other users in the network that share the same pilot sequence as that of user $jk$, i.e.,  $\pi_{jk}=\{(l,i): {\boldsymbol\varphi_{t_{li}}}={\boldsymbol\varphi_{t_{jk}}}\ \text{for}\ l=1,\cdots,L,\ \ k=1,\cdots,K, (l,i)\neq(j,k)\}$, and $\acute\pi_{jk}=\pi_{jk}\cup (j,k)$. The received uplink pilot signal matrix $\mathbf{Y}_j^p\in \mathbb{C}^{M\times\tau_\rho}$ at BS $j$ is:
\begin{equation}
	\label{Uplink_Training_Phase_eq1}
	\mathbf{Y}_j^p=\ \sum_{l=1}^{L}\sum_{k=1}^{K}{\sqrt{{\hat{p}}_{lk}}\mathbf{h}_{lk}^j\boldsymbol{\varphi}_{t_{lk}}^H}+\mathbf{N}_j^p,
\end{equation}
where matrix $\mathbf{N}_j^p\in\mathbb{C}^{M\times\tau_\rho}$ models the additive white Gaussian noise (AWGN), and its entries are i.i.d. and distributed as $\mathcal{N_C}(0,\,\sigma_{ul}^2)\,$ and $\sigma_{ul}^2$ is the noise power during uplink phase. BS $\textit{j}$ uses $\mathbf{Y}_j^p$ to estimate the channel vectors of all users in the network. Consider user $\textit{lk}$ with pilot sequence $\boldsymbol\varphi_{t_{lk}}$. To estimate the uplink channel $\mathbf{h}_{lk}^j$ from user $lk$, BS $\textit{j}$ first correlates $\mathbf{Y}_j^p$ with $\boldsymbol\varphi_{t_{lk}}$ to obtain vector $\mathbf{y}_{jlk}^p\in\mathbb{C}^{M\times1}$
\begin{equation}
	\label{Uplink_Training_Phase_eq2}
	 \mathbf{y}_{jlk}^p=\mathbf{Y}_j^p{\boldsymbol\varphi}_{t_{lk}}=\sum_{(n,i)\in\acute\pi_{lk}}{\sqrt{{\hat{p}}_{ni}}\mathbf{h}_{ni}^j{\boldsymbol\varphi}_{t_{ni}}^H}{\boldsymbol\varphi}_{t_{lk}}+\mathbf{N}_j^p{\boldsymbol\varphi}_{t_{lk}},
\end{equation}
where the noise vector $\mathbf{N}_j^p\boldsymbol{\varphi}_{t_{lk}}$ is distributed as 
$\mathcal{N_C}(\mathbf{0},\tau_\rho\sigma_{ul}^2\mathbf{I}_M)\,$. Using $\mathbf{y}_{jlk}^p$, next BS $\textit{j}$ finds the minimum mean square error (MMSE) estimate of $\mathbf{h}_{lk}^j$ \cite{kay1993fundamentals}, \cite{bjornson2017massive}:
\begin{equation}
	\label{Uplink_Training_Phase_eq3}
	{\hat{\mathbf{h}}}_{lk}^j=\mathbb{E}\{\mathbf{h}_{lk}^j|\mathbf{y}_{jlk}^p\}=\sqrt{\tau_\rho{\hat{p}}_{lk}}\mathbf{R}_{lk}^j(\boldsymbol{\psi}_{lk}^j)^{-1}\mathbf{y}_{jlk}^p,
\end{equation}
where $\boldsymbol{\psi}_{lk}^j =\sum_{(n,i)\in\acute\pi_{lk}}{\tau_\rho{\hat{p}}_{ni}}\mathbf{R}_{ni}^j+\sigma_{ul}^2\mathbf{I}_M$. Note that ${\hat{\mathbf{h}}}_{lk}^j$ is subject to pilot contamination from users who share the same pilot sequence ${\boldsymbol\varphi}_{t_{lk}}$. One can show that  ${\hat{\mathbf{h}}}_{lk}^j$ is distributed as ${\hat{\mathbf{h}}}_{lk}^j\sim \mathcal{N_C}(\mathbf{0},{\bf\Xi}_{lk,lk}^j)$, where ${\bf\Xi}_{jk,li}^j=\tau_p\sqrt{\hat{p}_{jk}\hat{p}_{li}}{\bf R}_{jk}^j({\mathbf\psi}_{jk}^j)^{-1}{\bf R}_{li}^j$ represents the channel estimation cross-covariance matrix for pilot-sharing users, given that $jk\neq li$. Invoking the orthogonality principle of the MMSE estimate, the estimation error ${\tilde{\mathbf{h}}}_{lk}^j=\ \mathbf{h}_{lk}^j - {\hat{\mathbf{h}}}_{lk}^j$ is independent of ${\hat{\mathbf{h}}}_{lk}^j$ and is distributed as ${\tilde{\mathbf{h}}}_{lk}^j\sim \mathcal{N_C}(\mathbf{0},\mathbf{C}_{lk}^j)$, where $\mathbf{C}_{lk}^j=\ \mathbf{R}_{lk}^j-\tau_\rho{\hat{p}}_{lk}\mathbf{R}_{lk}^j(\boldsymbol{\psi}_{lk}^j)^{-1}\mathbf{R}_{lk}^j$ is the estimation error covariance matrix. A remark on channel estimates of pilot sharing users follows.
\begin{rem}\label{rem:channel estimation}
Examining (\ref{Uplink_Training_Phase_eq3}), we find that pilot sharing users have the same $\mathbf{y}^p$ vectors and $\boldsymbol{\psi}$ matrices. Suppose $(m,r)\in\pi_{lk}$. We have ${\hat{\bf h}}_{mr}^l={\bf \Upsilon}_{mr}^{lk}{\hat{\bf h}}_{lk}^l$ where ${\bf \Upsilon}_{mr}^{lk}=\sqrt{\frac{{\hat p}_{mr}}{\hat p_{lk}}}{\bf R}_{mr}^l({\bf R}_{lk}^l)^{-1}$. Assuming that the spatial correlation matrices of these users are asymptotically linearly independent (as $M$ grows to infinity), the channel estimate vectors are not aligned (i.e., they are separable), and the channel estimation error covariance matrices do not take the form of scaled identity matrices. For uncorrelated fading model, however, where spatial correlation matrix is ${\bf{R}}_{lk}^j=\beta_{lk}^j\bf{I}$, the MMSE channel estimates of these users are aligned vectors that cannot be separated (so-called parallel channel estimates). Furthermore, the channel estimation error covariance matrices reduce to the scaled identity matrices.
\end{rem}

The assumption regarding the spatial correlation matrices mentioned in Remark \ref{rem:channel estimation} can be mathematically expressed as follows \cite{bjornson2017massive}:

$\bf{A 1:}$
$\liminf_M \frac{1}{M}\|{\bf{R}}_{jk}^j - \sum_{(l,i)\in\pi_{jk}} c_{li}{\bf{R}}_{li}^j\|_F^2>0$, $\forall j,k$\\
where $c_{li}\in\mathbb{R}$ $\forall l, i$. The asymptotic linear independence condition imposes greater restrictions than linear independence since it demands not only linear independence, but also stipulates that the subspace where the matrices differ must exhibit a norm that grows at a minimum linear rate with $M$ \cite{bjornson2017massive}. In Section \ref{sec:pilot assignment} we discuss how our proposed multi-cell PA scheme ensures that $\bf{A 1}$ is satisfied. 

\begin{figure*}
   \centering
   \begin{equation}
    \label{Uplink_Lower_Bound_eq2}
    {SINR}_{jk}^{ul}=\frac{p_{jk}{\bf v}_{jk}^H{\hat{\mathbf{h}}}_{jk}^j({\hat{\mathbf{h}}}_{jk}^j)^H\mathbf{v}_{jk}}{{\bf v}_{jk}^H(p_{jk}\mathbf{C}_{jk}^j+\sum\limits_{(l,i)\neq(j,k)}p_{li}({\hat{\mathbf{h}}}_{li}^j({\hat{\mathbf{h}}}_{li}^j)^H+\mathbf{C}_{li}^j)+\sigma_{ul}^2{\bf I}_M)\mathbf{v}_{jk}}.
    \end{equation}
    \rule{\textwidth}{0.4pt}
    \end{figure*}

\subsection{Uplink M-MMSE Detector}
During the uplink data transmission phase, each user transmits $\tau_u$ i.i.d. Gaussian data symbols. Let $s_{lk} \sim \mathcal{N_C}(0,p_{lk})$ denote a transmitted data symbol by user $lk$, and $p_{lk}$ be the corresponding data power. The received uplink data signal vector $\mathbf{y}_j\in\mathbb{C}^{M\times1}$ at BS $\textit{j}$ is:
\begin{equation}
	\label{Uplink_Data_Transmission_Phase_eq1}
	\mathbf{y}_j=\sum_{l=1}^{L}\sum_{k=1}^{K}{\mathbf{h}_{lk}^js_{lk}+}\mathbf{n}_j,
\end{equation}
where $\mathbf{n}_j\sim \mathcal{N_C}(\mathbf{0},\sigma_{ul}^2\mathbf{I}_M)$ models the AWGN. We denote the linear detector used by BS $j$ to detect the data signal of an arbitrary user $k$ in its cell as $\mathbf{v}_{jk}\in\mathbb{C}^{M\times1}$. Then the detected signal $\hat{s}_{jk}=\mathbf{v}_{jk}^H\mathbf{y}_j$ can be written as
\begin{equation}
	\begin{split}	
	\label{Uplink_Data_Transmission_Phase_eq2}
	\hat{s}_{jk}={\mathbf{v}_{jk}^H}{\hat{\mathbf{h}}}_{jk}^j{s}_{jk}+{\mathbf{v}_{jk}^H}{\tilde{\mathbf{h}}}_{jk}^j{s}_{jk} + \mathbf{v}_{jk}^H\sum_{(l,i)\neq(j,k)}{\mathbf{h}_{li}^j{s}_{li}}+ {\mathbf{v}_{jk}^H\mathbf{n}}_j	
\end{split}
\end{equation}
where the first term in (\ref{Uplink_Data_Transmission_Phase_eq2}) corresponds to the desired signal, whereas the second, the third, and the forth terms in (\ref{Uplink_Data_Transmission_Phase_eq2}) correspond to channel uncertainty, interference, and additive noise, and thus can be treated as noise in the signal detection. The achievable  ergodic SE of this user (measured in bits/sec/Hz) during uplink phase is lower bounded by \cite{bjornson2017massive}
\begin{equation}
	\label{Uplink_Lower_Bound_eq1}
	{\rm SE}_{jk}^{ul}=(1-\frac{\tau_\rho+\tau_d}{\tau_c})\mathbb{E}\{\log_2{(1+{SINR}_{jk}^{ul})}\},
\end{equation}
where the expectation is with respect to all channel estimates obtained at BS $j$, and the instantaneous effective SINR is given in (\ref{Uplink_Lower_Bound_eq2}). Note that ${SINR}_{jk}^{ul}$ in (\ref{Uplink_Lower_Bound_eq2}) is the form of a generalized Rayleigh quotient. Let $\mathbf{v}_{jk}^{M-MMSE}$ represent the optimal linear detector obtained from maximizing the Rayleigh quotient \cite{bjornson2017massive}, given all channel estimates at BS $j$. We have:
\begin{equation}
	\label{Uplink_Lower_Bound_eq3}
	\mathbf{v}_{jk}^{M-MMSE}={\boldsymbol{\Sigma}}_j{\hat{\mathbf{h}}}_{jk}^j
\end{equation}
where ${\boldsymbol{\Sigma}}_j=\left({\hat{\boldsymbol{\mathcal{H}}}}_j\mathbf{P}{{\hat{\boldsymbol{\mathcal{H}}}}_j}^H+\sum_{l=1}^{L}\sum_{i=1}^{K}{p_{li}\mathbf{C}_{li}^j}+\sigma_{ul}^2\mathbf{I}_M\right)^{-1}$, ${\hat{\boldsymbol{\mathcal{H}}}}_j=[{\hat{\mathbf{H}}}_{j1},{\hat{\mathbf{H}}}_{j2},\ldots,\ {\hat{\mathbf{H}}}_{jL}]\in\mathbb{C}^{M\times L K}$, $\mathbf{P}=diag(\mathbf{P}_1,\mathbf{P}_2,...,\mathbf{P}_L)\in\mathbb{R}^{LK\times L K}$, ${\hat{\mathbf{H}}}_{jl}=[{\hat{\mathbf{h}}}_{l1}^j,\cdots,{\hat{\mathbf{h}}}_{lK}^j]\in\mathbb{C}^{M\times K}$, and $\mathbf{P}_j=diag(p_{j1},\cdots,p_{jK})\in\mathbb{R}^{K\times K}$. We refer to $\mathbf{v}_{jk}^{M-MMSE}$ in (\ref{Uplink_Lower_Bound_eq3}) as M-MMSE detector.  One can show that this detector also minimizes the data detection error ${\rm MSE}_{jk}= \mathbb{E}\{|s_{jk}-\hat{s}_{jk}|^2|\{{\hat{\boldsymbol{\mathcal{H}}_j}}\}_{j=1}^L\}$\cite{bjornson2017massive}.
For uncorrelated fading model, where ${\bf{R}}_{jk}^l=\beta_{jk}^l \bf{I}$, then $\mathbf{v}_{jk}^{M-MMSE}$ in (\ref{Uplink_Lower_Bound_eq3}) reduces to the M-MMSE detector in \cite{li2017massive}. A remark on M-MMSE detectors of pilot sharing users in a cell follows.
\begin{rem} \label{rem: parallel detectors}	
Consider the term ${\hat{\boldsymbol{\mathcal{H}}}}_j\mathbf{P}{{\hat{\boldsymbol{\mathcal{H}}}}_j}^H$ in (\ref{Uplink_Lower_Bound_eq3}). where $\tau_p\leq rank({\hat{\boldsymbol{\mathcal{H}}}}_j\mathbf{P}{{\hat{\boldsymbol{\mathcal{H}}}}_j}^H)\leq LK$. For uncorrelated fading model, due to parallel channel estimates, $rank({\hat{\boldsymbol{\mathcal{H}}}}_j\mathbf{P}{{\hat{\boldsymbol{\mathcal{H}}}}_j}^H)$ reduces to its minimum $\tau_p$. Combined with the fact that the channel estimation error covariance matrices are the scaled identity matrices, we  note that M-MMSE detectors of pilot sharing users are aligned vectors that
cannot be separated from each other. This inseparability impedes the detector's efficiency in mitigating pilot contamination. On the other hand, for correlated fading model, assuming $\bf{A1}$, $rank({\hat{\boldsymbol{\mathcal{H}}}}_j\mathbf{P}{{\hat{\boldsymbol{\mathcal{H}}}}_j}^H)$ exceeds $\tau_p$ and approaches its maximum $LK$. Further, the channel estimation error covariance matrices are not the scaled identity matrices. Therefore, ${\hat{\boldsymbol{\mathcal{H}}}}_j\mathbf{P}{{\hat{\boldsymbol{\mathcal{H}}}}_j}^H+\sum_{l=1}^{L}\sum_{i=1}^{K}{p_{li}\mathbf{C}_{li}^j}$ is more diverse, in terms of the number of distinct eigenvalues and their multiplicity, which makes the inverted matrix in (\ref{Uplink_Lower_Bound_eq3}) a better linear transform for channel estimates. Different from uncorrelated fading model, M-MMSE detectors of pilot sharing users can be separable vectors. This separability enables the detectors to effectively mitigate interference.
\end{rem}
For comparison, let consider the S-MMSE detector in \cite{hoydis2013massive,bjornson2017massive}, denoted as $\mathbf{v}_{jk}^{\text{S-MMSE}}$ as state-of-the-art single-cell scheme, in which BS $j$ only knows the channel estimates of users in its cell. Expressing  $\mathbf{v}_{jk}^{\text{S-MMSE}}$ using our notations in , we have:
    
\begin{equation}
	\label{Uplink_Lower_Bound_eq4}
	\resizebox{0.5\textwidth}{!}{$
		{\mathbf{v}}_{jk}^{\text{S-MMSE}} = \underbrace{\left({\hat{\mathbf{H}}}_{jj}{\mathbf{P}}_j{\hat{\mathbf{H}}}_{jj}^H + \sum\limits_{i=1}^{K}{p_{ji}\mathbf{C}}_{ji}^j + \sum\limits_{l\neq j,\ l=1}^{L}\sum\limits_{i=1}^{K}{p_{li}\mathbf{R}}_{li}^j + \sigma_{ul}^2{\mathbf{I}_M}_{=\boldsymbol{\tilde\Sigma}_j}\right)^{-1}}{\hat{\mathbf{h}}}_{jk}^j
		$}\nonumber
\end{equation}
\subsection{Downlink M-MMSE Precoder}
During the downlink data transmission phase, each BS transmits $\tau_d$ i.i.d. Gaussian data symbols. Let $x_{lk} \sim \mathcal{N_C}(0,\rho_{lk})$ denote a transmitted data symbol for user $lk$ and  $\rho_{lk}$ be the corresponding data power. The received downlink data signal vector $\mathbf{y}_{jk}\in\mathbb{C}^{M\times1}$ at user $\textit{k}$ in cell $\textit{j}$ is
\begin{equation}
\label{Downlink_Data_Transmission_Phase_eq1}
    \mathbf{y}_{jk}  =
		\sum_{l=1}^{L}(\mathbf{h}_{jk}^l)^H\sum_{i=1}^{K}\mathbf{w}_{li}x_{li}+\mathbf{n}_{jk}
\end{equation}
where $\mathbf{w}_{li}\in\mathbb{C}^{M\times1}$ is the precoder used by BS $\textit{l}$ for user $\textit{i}$ in its cell, and $\mathbf{n}_{jk}\sim \mathcal{N_C}(\mathbf{0},\sigma_{dl}^2\mathbf{I}_M)$ models the AWGN. The achievable ergodic SE of this user during downlink phase can be lower bounded by \cite{bjornson2017massive}:
\begin{equation}
	\label{Downlink Lower Bound Ergodic Capacity eq1}
	{SE}_{jk}^{dl}=(1-\frac{\tau_\rho+\tau_u}{\tau_c})\log_2{(1+{SINR}_{jk}^{dl})}
\end{equation}
where the effective SINR, denoted as ${SINR}_{jk}^{dl}$, is
\begin{equation}
	\begin{split}
	\label{Downlink Lower Bound Ergodic Capacity eq2}
	&{SINR}_{jk}^{dl} =\\ & \frac{\rho_{jk}|\mathbb{E}\{\mathbf{w}_{jk}^H\mathbf{h}_{jk}^j\}|^2}{\sum_{l=1}^{L}\sum_{i=1}^{K}{\rho_{li}\mathbb{E}\{|\mathbf{w}_{li}^H\mathbf{h}_{jk}^l|^2\}}-\rho_{jk}|\mathbb{E}\{\mathbf{w}_{jk}^H\mathbf{h}_{jk}^j\}|^2+\sigma_{dl}^2}
	\end{split}
\end{equation}
and the expectations are with respect to the channel realizations. Different from ${\rm SE}_{jk}^{ul}$ in (\ref{Uplink_Lower_Bound_eq1}) which only depends on the user's detector $\mathbf{v}_{jk}$, ${\rm SE}_{jk}^{dl}$ in (\ref{Downlink Lower Bound Ergodic Capacity eq1}) depends on precoders of all users in the network. A join optimization of precoders across cells seems impractical \cite{bjornson2017massive}. Recently, an uplink-downlink duality for massive MIMO systems was established which proves that, given any sets of detectors and uplink transmit powers, $SINR_{jk}^{ul}$ = $SINR_{jk}^{dl}$, and thus $SE_{jk}^{ul}=SE_{jk}^{dl}$ if
\begin{equation}
	\label{UL/DL duality for Deterministic Downlink Rate with M-MMSE eq1}
	\mathbf{w}_{jk}=\frac{\mathbf{v}_{j k}}{\sqrt{\mathbb{E}\{\|\mathbf{v}_{j k}\|^2\}}}
\end{equation}
and downlink powers are allocated by Theorem \ref{Theorem: U/D Duality Power Control}\cite{Bjornson2016MassiveMF}. Next, we derive a deterministic approximation for $SINR_{jk}^{ul}$.
\section{Asymptotic Analysis} \label{sec: Achievable Rates}
Since $SE_{jk}^{ul}$ in (\ref{Uplink_Lower_Bound_eq1}) is difficult to compute for a system with finite dimensions, similar to \cite{hoydis2013massive} we consider the large system limit, where $M$ and $K$ grow infinitely large, while keeping a finite ratio $M/K$. Hence,  all vectors and matrices in this section should be perceived as sequences of vectors and matrices of growing dimensions. In the following, the notation $^{``}M\rightarrow\infty^{"}$ will refer to $M,K\rightarrow\infty$ such that $\frac{M}{K}\rightarrow c\in(1,\infty)$. We will derive a deterministic approximation of $SINR_{jk}^{ul}$ in (\ref{Uplink_Lower_Bound_eq2}), denoted as ${\overline{SINR}}_{jk}^{ul}$, for M-MMSE detector, such that $SINR_{jk}^{ul} -{\overline{SINR}}_{jk}^{ul}\stackrel{{\text{a.s.}}}{\longrightarrow}0$. This imply that $SE_{jk}^{ul}-\overline{SE}_{jk}^{ul}{\longrightarrow}0$, where $\overline{SE}_{jk}^{ul}=(1-\frac{\tau_\rho+\tau_d}{\tau_c})\log_2{(1+{\overline{SINR}}_{jk}^{ul})}$. The interpretation of these results is that, for given $M$ and $K$, $SINR$ approximation
and the corresponding rate become increasingly tight as $M$ and $K$ grow.
 
To enable the asymptotic analysis, we assume the spatial correlation matrices satisfy the following assumptions:

$\bf{A 2:}$ $\lim\sup_{M}\|{\bf R}_{lk}^j\|_2<\infty, \forall j,l,k$

$\bf{A 3:}$ $\lim\inf_{M}\frac{1}{M}tr({\bf R}_{lk}^j)>0, \forall j,l,k$

Before we continue, we recall two useful theorems from random matrix theory. \cite{hoydis2013massive}, \cite{hoydis2012random}, modifying them to align with our notations.
\begin{thm} \label{theorem: Asymptotic Analysis one}
Let ${\bf D}\in\mathbb{C}^{M\times M}$ and ${\mathbf S}_j\in\mathbb{C}^{M\times M}$ be Hermitian and positive semidefinite and let ${\hat{\boldsymbol{\mathcal{H}}}}_j$ be random with independent column vectors $p_{jk}^{\frac{1}{2}}\hat{\mathbf h}_{lk}^j\sim \mathcal{N_C}(\mathbf{0},\frac{1}{M}\hat{\mathbf R}_{lk}^j)$$\forall j,l,k$. Assume that $\bf{D}$ and $\hat{\mathbf R}_{lk}^j$ $\forall j,l,k$ have uniform bounded spectral norm (with respect to $M$). Then, for any $\rho>0$ we have:
\begin{equation}
	\label{Asymptotic_Analysis_eq1}
	\frac{1}{M}tr\left(\mathbf{D}\left({\hat{\boldsymbol{\mathcal{H}}}}_j\mathbf{P}{\hat{\boldsymbol{\mathcal{H}}}}_j^H+\mathbf{S}_j+\rho\mathbf{I}_M\right)^{-1}\right)-\frac{1}{M}tr(\mathbf{DT}_j(\rho))\stackrel{\text{a.s.}}{\to 0}\nonumber
\end{equation}
where $\mathbf{T}_j(\rho)\in\mathbb{C}^{M\times M}$ is 
\begin{equation}
	\label{Asymptotic_Analysis_eq2}
	\mathbf{T}_j\left(\rho\right)=\left(\frac{1}{M}\sum_{l=1}^L\sum_{k=1}^{K}\frac{\hat{\mathbf R}_{lk}^j}{1+\xi_{e_{lk}}^j\left(\rho\right)}+\mathbf{S}_j+\rho\mathbf{I}_M\right)^{-1}
\end{equation}
and $\xi_{e_{lk}}^j(\rho)=\lim_{t\rightarrow\infty}{\xi_{e_{lk}}^j}^{(t)}(\rho)
$ for $t = 1, 2,\cdots$, and
\begin{align}
	\label{Asymptotic_Analysis_eq3}
	&{\xi_{e_{lk}}^j}^{(t)}(\rho)=
    \\&\frac{1}{M}tr\left(\hat{\mathbf R}_{lk}^j\left(\frac{1}{M}\sum_{n=1}^L\sum_{i=1}^{K}\frac{\hat{\mathbf R}_{ni}^j}{1+{\xi_{e_{ni}}^j}^{(t-1)}(\rho)}+\mathbf{S}_j+\rho\mathbf{I}_M\right)^{-1}\right)\nonumber
\end{align}
	with initial values ${\xi_{e_{lk}}^j}^{(0)}(\rho)=\frac{1}{\rho} \forall j,l,k$. The one-to-one function $e_{lk}\in\{1,\cdots,LK\}$ is defined to map a user indexed by double indices $l, k$ into a user with single index $e_{lk}$ such that $e_{11}=1$ and $e_{LK}=LK$.
\end{thm}
\begin{thm} \label{theorem: Asymptotic Analysis two}
	Let $\mathbf{\Theta}\in\mathbb{C}^{M\times M}$ be Hermitian and positive semidefinite with uniformly bounded spectral norm (with respect to $M$). Under the conditions of Theorem \ref{theorem: Asymptotic Analysis one}, for any $\rho>0$ we have:
	\begin{equation}
		\begin{split}
			\label{Asymptotic_Analysis_eq4}
			\frac{1}{M}&tr(\mathbf{D}({\hat{\boldsymbol{\mathcal{H}}}}_j\mathbf{P}{\hat{\boldsymbol{\mathcal{H}}}}_j^H+\mathbf{S}_j+\rho\mathbf{I}_M)^{-1}\mathbf{\Theta}({\hat{\boldsymbol{\mathcal{H}}}}_j\mathbf{P}{\hat{\boldsymbol{\mathcal{H}}}}_j^H+\mathbf{S}_j+\rho\mathbf{I}_M)^{-1}) \\& -\frac{1}{M}tr(\mathbf{D}\mathbf{T}_{j}^\prime(\rho))\stackrel{\text{a.s.}}{\to 0}\nonumber
		\end{split}		
	\end{equation}
where $\mathbf{T}_{j}^\prime(\rho)\in\mathbb{C}^{M\times M}$ is
\begin{align}
\label{Asymptotic_Analysis_eq5}
\mathbf{T}_{j}^\prime(\rho) &= \mathbf{T}_j(\rho)\mathbf{\Theta} \mathbf{T}_j(\rho) \nonumber\\
&\quad + \mathbf{T}_j(\rho)\frac{1}{M}\sum_{n=1}^L\sum_{i=1}^{K}\frac{\hat{\mathbf R}_{ni}^j \acute{\xi}_{e_{ni}}^{jk}(\rho)}{\left(1+\xi_{e_{ni}}^j(\rho)\right)^2}\mathbf{T}_j(\rho)
\end{align}

Also, $\mathbf{T}_j(\rho)$ and ${\xi}_{e_{lk}}^j(\rho) \forall j,l,k$ are given by Theorem~\ref{theorem: Asymptotic Analysis one}, and $\boldsymbol{\acute\xi}^{jk}(\rho)=[{\acute\xi}^{jk}_{1}(\rho),\cdots,{\acute\xi}^{jk}_{LK}(\rho)]^T$ is calculated as
	\begin{equation}
		\label{Asymptotic_Analysis_eq6}
		\boldsymbol{\acute\xi}^{jk}(\rho)=\left(\mathbf{I}_{LK}-\mathbf{J}_j(\rho)\right)^{-1}\boldsymbol{v}_{jk}(\rho)\nonumber
	\end{equation}
for $1\le k,i\le K, 1\le j,l,n\le L$, the entries of matrix $\mathbf{J}_j(\rho)\in\mathbb{C}^{LK\times LK}$ and vector $\boldsymbol{v}_{j}(\rho)\in\mathbb{C}^{LK}$ are defined as
	\begin{equation}
		\label{Asymptotic_Analysis_eq7}
		[\mathbf{J}_j(\rho)]_{e_{lk}e_{ni}}=\frac{\frac{1}{M}tr\left(\hat{\mathbf R}_{lk}^j\mathbf{T}_j(\rho)\hat{\mathbf R}_{ni}^j\mathbf{T}_j(\rho)\right)}{M\left(1+\xi_{e_{ni}}^j(\rho)\right)^2}\nonumber
	\end{equation}
	\begin{equation}
		\label{Asymptotic_Analysis_eq8}
		[\boldsymbol{v}_{j}(\rho)]_{e_{ni}}=\frac{1}{M}tr\left(\hat{\mathbf R}_{ni}^j\mathbf{T}_j(\rho)\mathbf{\Theta}{\mathbf T}_j(\rho)\right)\nonumber
	\end{equation}
\end{thm}
In Theorem \ref{theorem: Asymptotic Analysis three} we derive the deterministic SINR approximation ${\overline{SINR}}_{jk}^{ul}$, which is averaged over small scale fading and depends only on spatial correlation matrices ${\bf{R}}_{lk}^j \, \forall l,k,j$.
\begin{thm} \label{theorem: Asymptotic Analysis three} Assume that ${\bf A2}$ and ${\bf A3}$ are hold. Then ${SINR}_{jk}^{ul}-{\overline{SINR}}_{jk}^{ul}\underset{{M\rightarrow\infty}}{{\buildrel a.s. \over \longrightarrow }}0$, where ${\overline{SINR}}_{jk}^{ul}$ is 
\begin{equation}
	\label{Asymptotic_Analysis_eq9}
	{\overline{SINR}}_{jk}^{ul}=\frac{p_{jk}\delta_{jk}^2}{\sum_{(l,i)\in\pi_{jk}}{p_{li}\delta_{jlik}^2}+\sum\limits_{(l,i)\notin\acute\pi_{jk}}{{p_{li}}\mu_{jlik}}+\delta_{jk}^{\prime\prime}\sigma_{ul}^2}
\end{equation}
where all $\delta_{jk}$, $\delta_{jlik}$, $\mu_{jlik}$, and $\delta_{jk}^{\prime\prime}\forall j,l,i,k$ are defined in appendix \ref{Appendix:app B}, and
\begin{enumerate}
\item$\mathbf{S}_j=\frac{1}{M}\sum_{l=1}^{L}\sum_{i=1}^{K}{p_{li}\mathbf{C}_{li}^j}$, $\mathbf{T}_j(\rho)$ is obtained from (\ref{Asymptotic_Analysis_eq2}) by replacing $\rho=\frac{1}{M}\sigma_{ul}^2$.

\item${\mathbf{T}^\prime}_{j}(\rho)$ is obtained from (\ref{Asymptotic_Analysis_eq5}) by replacing $\boldsymbol{\Theta}$ with channel estimate cross-covariance matrices of  pilot-sharing users specified in the appendix and $\rho=\frac{1}{M}\sigma_{ul}^2$.

\item$\mathbf{T}_j^{\prime\prime}(\rho)$ is obtained from (\ref{Asymptotic_Analysis_eq5}) by replacing $\mathbf{\Theta}=\mathbf{I}_M$, and $\rho=\frac{1}{M}\sigma_{ul}^2$.
\end{enumerate}
\end{thm}
{\bf{Proof}}. see appendix \ref{Appendix:app B}.
\begin{rem}\label{rem:APPENDIX}
	To highlight the differences between our results and the large-scale uplink SINR approximations presented in \cite{li2017massive} and \cite{hoydis2013massive}, which respectively considered M-MMSE and S-MMSE detectors, it is important to note that the simplifications in these works stem from the specific assumptions they employed. In \cite{li2017massive}, the use of an uncorrelated channel model allows for a \textit{single application} of Lemma 2 to $\boldsymbol{\Sigma}_j$, effectively removing  ${\bf\hat{h}}_{jk}^j$  from $\boldsymbol{\Sigma}_j$ before applying Theorems 1 and 2. Similarly, in \cite{hoydis2013massive}, the prohibition of pilot reuse within each cell permits a \textit{single application} of Lemma 2 to $\boldsymbol{\tilde\Sigma}_j$, which suffices to eliminate  ${\bf\hat{h}}_{jk}^j$  from $\boldsymbol{\tilde\Sigma}_j$ prior to utilizing Theorems 1 and 2. In contrast, our derivation involves a correlated channel model, requiring a more intricate approach. Specifically, we must apply Lemma 2 to $\boldsymbol{\Sigma}_j$ \textit{multiple times}, not only to remove  ${\bf\hat{h}}_{jk}^j$  but also to eliminate all channel estimates corresponding to users sharing the same pilot as user $jk$ from $\boldsymbol{\Sigma}_j$. This necessity for repeated application of Lemma 2 significantly complicates the derivation process.
\end{rem}
We note that the numerator of ${\overline{SINR}}_{jk}^{ul}$ in (\ref{Asymptotic_Analysis_eq9}) corresponds to the desired signal. Considering the denumerator of ${\overline{SINR}}_{jk}^{ul}$, the first term is known as the coherent interference (a.k.a. pilot contamination ) caused by other users with the same pilot. The second and the third term correspond to the non-coherent interference and AWGN, respectively. As $M$ grows to infinity, the second and the third terms can be significantly reduced. However, the first term cannot be diminished and depends on the PA scheme. It is well known that the pilot contamination limits the capacity of a massive MIMO system \cite{bjornson2017massive}. This motivates our multi-cell PA scheme in Section \ref{sec:pilot assignment}, which aims at minimizing the coherent interference power.

\section{Pilot Contamination and PA}
\label{sec:pilot assignment}
\label{sec: Multi-Cell pilot assignment algorithms}
To motivate our multi-cell PA scheme let start with the single-cell PA scheme proposed in \cite{zhu2016weighted}. For uncorrelated fading model where spatial correlation matrix reduces to ${\bf{R}}_{lk}^j=\beta_{lk}^j\bf{I}$, the authors in \cite{zhu2016weighted} proposed a scheme (based on weighted graph coloring) aiming at mitigating pilot contamination, and showed that it outperforms the scheme in \cite{zhu2015smart}. To obtain the scheme in \cite{zhu2016weighted}, the authors considered maximizing the sum of average uplink rate of users, where the instantaneous effective SINR is approximated using large scale fading coefficients. 
 Given each BS uses matched filter detector and $M\rightarrow\infty$, \cite{zhu2016weighted} found the following deterministic SINR approximation given uniform power allocation:
\begin{equation}
	\label{Single-Cell pilot assignment algorithms eq2}
	{SINR}_{jk}^{ul}\approx\frac{{\beta_{jk}^j}^2}{\sum_{\left(l,i\right)\in\pi_{jk}}{\beta_{li}^j}^2}
\end{equation}
where the numerator and the denumerator correspond to the desired signal and the coherent interference, respectively. Although (\ref{Single-Cell pilot assignment algorithms eq2}) is derived for matched filter detector at BS, one can show that (\ref{Single-Cell pilot assignment algorithms eq2}) remains unchanged for M-MMSE detector. To construct the weighted graph, the authors in \cite{zhu2016weighted} defined the following metric $\zeta_{jk,li}$, which measures the potential pilot contamination strength between two pilot sharing users in neighboring cells, user $jk$ and user $li$,  $l \neq j$:
\begin{equation}
	\label{Single-Cell pilot assignment algorithms eq3}
	\zeta_{jk,li}=\left(\frac{\beta_{li}^j}{\beta_{jk}^j}\right)^2+\left(\frac{\beta_{jk}^l}{\beta_{li}^l}\right)^2
\end{equation}
To demonstrate the inefficiency of single-cell PA scheme in \cite{zhu2016weighted} on effective suppression of the pilot contamination when each BS uses M-MMSE detector, we consider a simple scenario where user $jk$ and user $li$ are sharing the same pilot. For uncorrelated fading model from (\ref{Uplink_Training_Phase_eq3}) we have:
\begin{equation}
	\label{Multi-Cell pilot assignment algorithms eq1}
	\frac{\beta_{li}^j}{\beta_{jk}^j}{\hat{\mathbf{h}}}_{jk}^j={\hat{\mathbf{h}}}_{li}^j \qquad k,i=1,\ldots,K \quad   l\neq\ j
\end{equation}
indicating that the channel estimates of these users are aligned vectors. Considering the metric in (\ref{Single-Cell pilot assignment algorithms eq3}) we recognize that the ratio $\left(\frac{\beta_{li}^j}{\beta_{jk}^j}\right)^2$ determines the pilot contamination strength imposed on users in cell $j$. Since the scheme in \cite{zhu2016weighted} targets minimizing the pilot contamination, this ratio should be minimized. Applying this constraint to (\ref{Multi-Cell pilot assignment algorithms eq1}) we conclude that a good quality channel estimate ${\hat{\mathbf{h}}}_{jk}^j$ with a small $tr({\bf{C}}_{jk}^j)$ corresponds to a bad quality channel estimate  ${\hat{\mathbf{h}}}_{li}^j$ with a much larger $tr({\bf{C}}_{li}^j)$. Let examine how this conclusion impacts M-MMSE detector in (\ref{Uplink_Lower_Bound_eq3}). Considering the matrix inverse in (\ref{Uplink_Lower_Bound_eq3}), we note that the channel estimation error for $(L-1)K$ users outside cell $j$ are much larger than the ones for $K$ users in cell $j$, since they are sharing pilots with users inside cell $j$, rendering the M-MMSE detector less effective in suppressing the interference.

\subsection{Optimality Condition for Our Multi-Cell PA Scheme}
\label{Our Proposed Multi-Cell pilot assignment Scheme}
We consider the coherent interference in (\ref{Asymptotic_Analysis_eq9}), denoted as $\eta_{jk}$. Noting that for designing PA scheme, power allocation is irrelevant, we assume uniform (pilot and data) power allocation. Therefore:\begin{equation}
	\begin{split}
		\label{Multi-Cell pilot assignment algorithms eq3}
		\eta_{jk}={\sum_{(l,i)\in\pi_{jk}}\delta_{jlik}^2}
	\end{split}
\end{equation}
\begin{thm} \label{theorem: Multi-Cell pilot assignment algorithms} Consider user $jk$, where 
	$\pi_{jk}$ is the set of users in the network that share the same pilot sequence as that of user $jk$.
	The pilot contamination imposed on this user is zero, i.e., $\eta_{jk}=0$, under the following condition:
\begin{equation}
	\label{Multi-Cell pilot assignment algorithms eq4}
	 tr\left({\bf R}_{jk}^j{\bf R}_{li}^j\right)=0\quad for\,\, (l,i) \in \pi_{jk}
\end{equation}
\end{thm}
Our proof of Theorem \ref{theorem: Multi-Cell pilot assignment algorithms} is based on the assumption that each BS is equipped with a uniform linear array (ULA) with half wavelength antenna spacing\footnote{The proposed multi-cell PA schemes are not limited to ULAs and apply equally to uniform planar arrays (UPAs) or other array geometries, as they rely on spatial correlation matrices that capture angular characteristics independent of the array structure. Extending to UPAs can further improve SE by leveraging elevation-domain resolution.}. Hence we have the following multipath model ${\bf h}_{lk}^j = \sum_{n\in P_{jlk}} g_{{jlk}_n}{\bf a}(\phi_{{jlk}_n})$,
where $P_{jlk}$ is a set of all possible i.i.d. paths, $g_{{jlk}_n}$ and $\phi_{{jlk}_n}$ are, respectively, the random complex gain and the random angle of arrival (AOA) of the $n$-th path between user $lk$ and BS $j$. We assume $g_{{jlk}_n}$ is independent over BS indices $j,l$, user index $k$, path index $n$, is zero mean and has variance of ${\mathbb E}\{|g_{{jlk}_n}|^2\}$. 
The steering vector is:
\begin{equation}
	\label{Multi-Cell pilot assignment algorithms eq6}
	{\bf a}(\phi_{{jlk}_n})=[\begin{array}{llll}
		1 & e^{-j 2 \pi cos(\phi_{{jlk}_n})} & \cdots & e^{-j 2 \pi cos(\phi_{{jlk}_n})(M-1)}\nonumber
	\end{array}]^T
\end{equation}
and steering vectors corresponding to distinct AOAs are asymptotically orthogonal (as $M \rightarrow \infty$).
 From the channel model we obtain:
\begin{equation}
	\begin{split}
	\label{Spatial correlation matrix}
	{\bf R}_{lk}^j=\mathbb{E}\{{\bf h}_{lk}^j{{\bf h}_{lk}^j}^H\}=\sum_{n\in P_{jlk}} {\mathbb E}\{|g_{{jlk}_n}|^2\} {\mathbb E}\{\underbrace{{\bf a}(\phi_{{jlk}_n}){\bf a}(\phi_{{jlk}_n})^H}_{\text{\bf A}}\}
	\end{split}
\end{equation}

where the $(m,m^\prime)$-th entry of matrix $\bf A$ is $[{\bf A}]_{m,m^\prime}=e^{j2\pi(m-m^\prime)cos(\phi_{jlk_n})}$. Since ${\bf A}$ has rank one, (\ref{Spatial correlation matrix}) means that each path adds an eigenvalue to ${\bf R}_{lk}^j$. 
Therefore, ${\bf R}_{lk}^j=span\{{\bf{a}}(\phi_{{jlk}_n}),\ cos(\phi_{{jlk}_n})\in[b_1\ ,b_2]\ ,b_1<b_2\in[-1\ ,1],\,\text{and}\, {n}\in P_{jlk}\}$, and we can approximate the eignevalue decomposition (EVD) of ${\bf R}_{lk}^j$ as the following:
\begin{equation}
	\label{Multi-Cell pilot assignment algorithms eq9}
	{\bf R}_{lk}^j\approx {\bf V}{\bf D}_{lk}^j{{\bf V}}^H
\end{equation}
where the $m$-th column of ${\bf V}$ is $\frac{1}{\sqrt M}{{\bf{a}}}(\phi)$ with $cos(\phi)=-1+\frac{2(m-1)}{M}$, $m=1,\cdots,M$, ${\bf D}_{lk}^j=diag(\lambda_{{jlk}_1},\cdots,\lambda_{{jlk}_M})$ is a diagonal matrix that contains eigenvalues of ${\bf R}_{lk}^j$ , and $\lambda_{{jlk}_m}$ corresponds to $m$-th column in ${\bf V}$. We will use the approximation in (\ref{Multi-Cell pilot assignment algorithms eq9}) to prove Theorem \ref{theorem: Multi-Cell pilot assignment algorithms}. The proof follows. 

{\bf{Proof}}. In Appendix \ref{Appendix:app B} we showed that $\delta_{jlik}$ in (\ref{Multi-Cell pilot assignment algorithms eq3}) is composed of the basic functions ${\tilde f}_{0}^{\pi_{jk}}({\bf\Xi}_{jk,jk}^j,{\bf I},\frac{1}{M}{{\bf T}_j},{\bf \Upsilon}_{li}^{jk})$ and ${\tilde f}_{0}^{\pi_{jk}}({\bf\Xi}_{jk,jk}^j,{\bf \Upsilon}_{mr}^{jk},\frac{1}{M}{{\bf T}_j},{\bf \Upsilon}_{li}^{jk})$. Hence, $\delta_{jlik}$ is zero if these basic functions are zero. Based on the EVD in (\ref{Multi-Cell pilot assignment algorithms eq9}) we rewrite ${\tilde f}_{0}^{\pi_{jk}}({\bf\Xi}_{jk,jk}^j,{\bf I},{{\bf T}_j},{\bf\Upsilon}_{li}^{jk})$ for $(l,i)\in\pi_{jk}$ as follows: 
\begin{equation}
	\label{Multi-Cell pilot assignment algorithms eq10}
	 tr\left({\bf V}{\bf D}_{jk}^j{\bf D}_{li}^j\left(\sum_{(l^\prime,i^\prime)\in\pi_{jk}}{p_{l^\prime i^\prime}\tau_\rho {\bf D}_{l^\prime i^\prime}^j}+\sigma_{ul}^2\mathbf{I}_M\right)^{\text{\footnotesize-1}}{\bf D}_{T}^j{\bf V}^H\right)
\end{equation}
where ${\bf D}_{T}^j$ is a diagonal matrix containing the eigenvalues of ${\bf T}_j(\rho)$. We note that (\ref{Multi-Cell pilot assignment algorithms eq10}) becomes zero when $\sum_{(l,i)\in\pi_{jk}} tr({\bf D}_{jk}^j{\bf D}_{li}^j)=
\sum_{(l,i)\in\pi_{jk}} tr({\bf R}_{jk}^j{\bf R}_{li}^j)=0$. Similarly, we can rewrite ${\tilde f}_{0}^{\pi_{jk}}({\bf\Xi}_{jk,jk}^j,{\bf \Upsilon}_{mr}^{jk},\frac{1}{M}{{\bf T}_j},{\bf \Upsilon}_{li}^{jk})$ and show
that it becomes zero under the same condition. This condition implies that all terms of $\eta_{jk}$ in (\ref{Multi-Cell pilot assignment algorithms eq3}) are zero. Thus users in $\acute\pi_{jk}$ appear to have orthogonal pilots, even though they share the same pilot.
\begin{rem}\label{rem:multy-cell PA}
Interestingly, when the condition in (\ref{Multi-Cell pilot assignment algorithms eq4}) holds true, the desired signal power in (\ref{Asymptotic_Analysis_eq9}) is maximized. This is because from (\ref{APPENDIX B: eq6xxx5}) in appendix we have $\delta_{jk}={\tilde f}_{N}^{\pi_{jk}}({\bf\Xi}_{jk,jk}^j,{\bf I},\frac{1}{M}{{\bf T}_j},{\bf I})$. When the condition in (\ref{Multi-Cell pilot assignment algorithms eq4}) is satisfied,  the first term in $\delta_{jk}$, which is positive, remains unchanged, while the remaining terms inside the sum that are positive become zero.
\end{rem}
We have noted before that for correlated fading model the MMSE channel estimates and the M-MMSE detectors are not aligned (are separable) when ${\bf A1}$ is satisfied. Expanding the term $\|{\bf{R}}_{jk}^j - \sum_{(l,i)\in\pi_{jk}} c_{li}{\bf{R}}_{li}^j\|_F^2$ we obtain:
\begin{equation}
	\begin{split}
		\label{asymptotically linearly independen prove}
		tr({\bf{R}}_{jk}^j)^2+tr\left(\sum_{(l,i)\in\pi_{jk}} c_{li}{\bf{R}}_{li}^j\right)^2-2tr\left(\sum_{(l,i)\in\pi_{jk}} c_{li}{\bf{R}}_{jk}^j{\bf{R}}_{li}^j\right)\nonumber
	\end{split}
\end{equation}
This implies that when the sufficient condition in (\ref{Multi-Cell pilot assignment algorithms eq4}) holds true, ${\bf A1}$ is satisfied. For a single-cell system where pilot reuse within the cell is allowed, the authors in \cite{you2015pilot} proved that the minimum of the sum of mean square errors of channel estimates is achieved when the spatial correlation matrices of any two pilot sharing users are orthogonal, i.e.,
\begin{equation}
	\label{Multi-Cell pilot assignment algorithms eq14}
	\theta({\bf R}_{jk}^j,{\bf R}_{li}^j)=\arccos\frac{tr\left({\bf R}_{jk}^j{\bf R}_{li}^j\right)}{\|{\bf R}_{jk}^j\|_F\|{\bf R}_{li}^j\|_F}=90^{\circ}, \forall (l,i) \in \pi_{jk}
\end{equation}
While conditions (\ref{Multi-Cell pilot assignment algorithms eq4}) and (\ref{Multi-Cell pilot assignment algorithms eq14}) may appear similar, they are, in fact, distinct. Condition (\ref{Multi-Cell pilot assignment algorithms eq14}) is (\ref{Multi-Cell pilot assignment algorithms eq4}) divided by {\footnotesize$\|{\bf R}_{jk}^j\|_F=\sqrt{tr({\bf R}_{jk}^j{{\bf R}_{jk}^j}^H)}=\sqrt{\sum_{n\in P_{jjk}}{\lambda_{{jjk}_n}^2}}$} and {\footnotesize$\|{\bf R}_{li}^j\|_F=\sqrt{tr({\bf R}_{li}^j{{\bf R}_{li}^j}^H)}=\sqrt{\sum_{n\in P_{jli}}{\lambda_{{jli}_n}^2}}$}.
This division reduces the eigenvalue effects and concentrates more on orthogonal support of eigenvector. 
While condition (\ref{Multi-Cell pilot assignment algorithms eq4}) exactly minimizes the interference power of users who reuse the same pilot, condition (\ref{Multi-Cell pilot assignment algorithms eq14}) minimizes the overlapping AOA intervals of users who reuse the same pilot, which only partially contributes to interference power minimization.

\subsection{Our Proposed Multi-Cell PA Scheme}
Algorithm \ref{alg:alg1} outlines the pseudocode of our proposed multi-cell PA  scheme, in which we leverage Theorem \ref{theorem: Multi-Cell pilot assignment algorithms} in two crucial steps. First, we identify users producing higher interference power when using the same pilot, enabling us to strategically assign them orthogonal pilots. Second, we determine users producing lower interference power when using the same pilot, enabling us to assign them the same pilot. Each BS can run Algorithm \ref{alg:alg1} distributively, assuming that it has the values of $tr({\bf R}_{jk}^j {\bf R}_{li}^j)$ for all $k,l,i,j$. This becomes possible when BS $j$ sends the values of $tr({\bf R}_{jk}^j {\bf R}_{li}^j)$ for all $k,l,i$ to other BSs, requiring BS $j$ to send total of $\frac{LK}{2}(LK-1)$ real values, which it is independent of $M$ and remains practical for moderate $L$ and $K$.

\begin{algorithm}[H]
	\caption{Our proposed multi-cell PA algorithm.}\label{alg:alg1}
	\begin{algorithmic}[1]
		\REQUIRE The spatial correlation matrices $\mathbf{R}$, the network pilot matrix $\bf\Phi$, the user set ${\mathcal S}=\{(l,k), \forall l,k\}$, and orthogonal pilot length $\eta=\tau_p$.
		\ENSURE $\pi_{lk}$ and $\acute\pi_{lk}$ for $\textit{l} = 1,\ldots,L$ and $\textit{k} = 1,\ldots,K$.
		\STATE \textbf{Initialization}: $\pi_{lk}=\emptyset$ and $\acute\pi_{lk}=(l,k)$ $\forall\textit{l}, \textit{k}$.
		\WHILE {$\eta\neq 0$ for $(l,k)$, $(\acute{l},\acute{k})\in\mathcal{S}$, and $(\acute{\textit{l}},\acute{\textit{k}})\neq(\textit{l},\textit{k})$}
		\STATE $(l,k),(\acute{l},\acute{k}) = \underset{(l,k),(\acute{\textit{l}},\acute{\textit{k}})}{\arg\max}\sum_{j=1}^L tr({\bf R}_{lk}^j{\bf R}_{\acute{\textit{l}}\acute{\textit{k}}}^j)$
		\STATE If $(\acute{l},\acute{k})\in\mathcal{S}$ update $\pi_{{\textit{l}}{\textit{k}}} \gets \pi_{{\textit{l}}{\textit{k}}}\cup (\acute{l},\acute{k})$,  $\acute\pi_{{\textit{l}}{\textit{k}}} \gets \acute\pi_{{\textit{l}}{\textit{k}}}\cup (\acute{l},\acute{k})$, ${\eta} \gets {\eta}- 1$ and ${\mathcal S} \gets {\mathcal S}	\setminus (\acute{l},\acute{k})$.
		\STATE If $(l,k)\in\mathcal{S}$ update $\pi_{{\acute{l}}{\acute{k}}} \gets \pi_{{\acute{l}}{\acute{k}}}\cup ({l},{k})$ and  $\acute\pi_{{\acute{l}}{\acute{k}}} \gets \acute\pi_{{\acute{l}}{\acute{k}}}\cup ({l},{k})$, ${\eta} \gets {\eta}- 1$ and ${\mathcal S} \gets {\mathcal S}	\setminus ({l},{k})$.		
		\ENDWHILE
		\WHILE {$(\acute{l},\acute{k})\notin\pi_{{l}{k}}$ for $(\textit{l},\textit{k})\notin{\mathcal S}$}
		\STATE $(\acute{\textit{l}},\acute{\textit{k}}) = \arg\min_{(\acute{\textit{l}},\acute{\textit{k}})}\sum_{j=1}^L tr({\bf R}_{lk}^j{\bf R}_{\acute{\textit{l}}\acute{\textit{k}}}^j)$		
		\STATE Update $\pi_{{\textit{l}}{\textit{k}}} \gets \pi_{{\textit{l}}{\textit{k}}}\cup (\acute{l},\acute{k})$ and $\acute\pi_{{\textit{l}}{\textit{k}}} \gets \acute\pi_{{\textit{l}}{\textit{k}}}\cup (\acute{l},\acute{k})$
		\STATE Update $\pi_{{\acute{l}}{\acute{k}}} = \pi_{{{l}}{{k}}}$ and  $\acute\pi_{{\acute{l}}{\acute{k}}} = \acute\pi_{{{l}}{{k}}}$
        \STATE Update ${\mathcal S} \gets {\mathcal S}	\setminus (\acute{l},\acute{k})$
		\ENDWHILE
	\end{algorithmic}
\end{algorithm}
\subsection{Our Proposed Scalable  Multi-Cell PA Scheme}
To address scalability as the number of cells $L$ increases, we propose a partial M-MMSE (P-MMSE) detector, where each BS $j$ relies only on the channel estimates of a selected subset of users ${\mathcal I}_j$, rather than all users. The subset ${\mathcal I}_j$ is defined by introducing a threshold $\gamma \in (0,1]$ on the large-scale fading coefficients ${\mathcal I}_j=\{(l,i) \ | \beta_{l,i}^j\geq\gamma\min\{\beta_{j,k}^j \forall k \}\}$. This selection ensures that only users with sufficiently strong channel gains to BS $j$ are included, reducing channel estimation complexity without significant performance loss. The P-MMSE detector vector for user $k$ in cell $j$ is given by:
\begin{equation}
    \label{P-MMSE eq}
        \mathbf{v}_{jk}^{P-MMSE} = \left( \sum_{(l,i)\in{\mathcal I}_j}  p_{li}\left({\hat{\mathbf{h}}}_{li}^j ({{\hat{\mathbf{h}}}_{li}^j})^H + \mathbf{C}_{li}^j\right) + \sigma_{ul}^2\mathbf{I}_M \right)^{-1} {\hat{\mathbf{h}}}_{jk}^j
\end{equation}
Here, $\gamma$ controls the trade-off between complexity and performance: $\gamma=0$ recovers the full M-MMSE, while $\gamma=1$ approaches S-MMSE. Since users from distant cells with weak large-scale fading contribute negligible interference, the size of ${\mathcal I}_j$ remains finite even as $L\to\infty$, ensuring scalability. Based on this approach, we further develop a scalable multi-cell PA scheme compatible with P-MMSE processing. In the distributed implementation, BS $j$ computes and shares only $tr({\bf R}_{li}^j{\bf R}_{\acute{l}\acute{i}}^j)$ for $(l,i)$,$(\acute{l},\acute{i}) \in {\mathcal I}_j$, significantly reducing information exchange overhead. The pseudocode for this scalable PA algorithm is provided in Algorithm \ref{alg:alg2}, leveraging the optimality condition from Theorem 4.
Table \ref{tab:pa_comparison} summarizes the information exchange overhead of the proposed multi-cell PA schemes and the single-cell PA scheme in \cite{zhu2016weighted}, highlighting that as $L\to\infty$, only the scalable multi-cell PA scheme maintains bounded overhead.
\begin{algorithm}[h]

	\caption{Our proposed scalable multi-cell PA algorithm.}\label{alg:alg2}
	\begin{algorithmic}[1]
		\REQUIRE The spatial correlation matrices $\mathbf{R}$, the network pilot matrix $\bf\Phi$, the user set ${\mathcal I}_j \forall j$, and orthogonal pilot length $\eta=\tau_p$ .
		\ENSURE $\pi_{lk}$ and $\acute\pi_{lk}$ for $\textit{l} = 1,\ldots,L$ and $\textit{k} = 1,\ldots,K$.
		\STATE \textbf{Initialization}: $\pi_{lk}=\emptyset$ and $\acute\pi_{lk}=(l,k)$ $\forall\textit{l}, \textit{k}$.
		\WHILE {$\eta\neq 0$ for $(\textit{l}, \textit{k}),(\acute{l},\acute{k})\in {\mathcal I}_j \forall j$, and $(\textit{l}, \textit{k})\neq(\acute{l},\acute{k})$}
		\STATE $(l,k),(\acute{l},\acute{k}) = \underset{(l,k),(\acute{\textit{l}},\acute{\textit{k}})}{\arg\max}\underset{(\textit{l}, \textit{k}),(\acute{l},\acute{k})\in {\mathcal I}_j}{\sum_{j=1}^L} tr({\bf R}_{lk}^j{\bf R}_{\acute{\textit{l}}\acute{\textit{k}}}^j)$
		\STATE If $(\acute{l},\acute{k})\in{\mathcal I}_j \forall j$ update $\pi_{{\textit{l}}{\textit{k}}} \gets \pi_{{\textit{l}}{\textit{k}}}\cup (\acute{l},\acute{k})$,  $\acute\pi_{{\textit{l}}{\textit{k}}} \gets \acute\pi_{{\textit{l}}{\textit{k}}}\cup (\acute{l},\acute{k})$, ${\eta} \gets {\eta}- 1$ and ${\mathcal I}_j \gets {\mathcal I}_j	\setminus (\acute{l},\acute{k}) \forall j$.
		\STATE If $(l,k)\in{\mathcal I}_j \forall j$ update $\pi_{{\acute{l}}{\acute{k}}} \gets \pi_{{\acute{l}}{\acute{k}}}\cup ({l},{k})$ and  $\acute\pi_{{\acute{l}}{\acute{k}}} \gets \acute\pi_{{\acute{l}}{\acute{k}}}\cup ({l},{k})$, ${\eta} \gets {\eta}- 1$ and ${\mathcal I}_j \gets {\mathcal I}_j	\setminus ({l},{k}) \forall j$.
		\ENDWHILE
		\WHILE {$(\acute{l},\acute{k})\notin\acute\pi_{{l}{k}}$ for $(\textit{l}, \textit{k})\notin {\mathcal I}_j \forall j$}
		\STATE $(\acute{\textit{l}},\acute{\textit{k}}) = \underset{{(\acute{\textit{l}},\acute{\textit{k}})}}{\arg\min}\ \underset{(\acute{l},\acute{k})\in {\mathcal I}_j}{\sum_{j=1}^L} tr({\bf R}_{lk}^j{\bf R}_{\acute{\textit{l}}\acute{\textit{k}}}^j)$
		
		\STATE Update $\pi_{{\textit{l}}{\textit{k}}} \gets \pi_{{\textit{l}}{\textit{k}}}\cup (\acute{l},\acute{k})$ and $\acute\pi_{{\textit{l}}{\textit{k}}} \gets \acute\pi_{{\textit{l}}{\textit{k}}}\cup (\acute{l},\acute{k})$
		\STATE Update $\pi_{{\acute{l}}{\acute{k}}} = \pi_{{{l}}{{k}}}$ and  $\acute\pi_{{\acute{l}}{\acute{k}}} = \acute\pi_{{{l}}{{k}}}$.
        \STATE Update ${\mathcal I}_j \gets {\mathcal I}_j	\setminus (\acute{l},\acute{k}) \ \forall j$
		\ENDWHILE
	\end{algorithmic}
    
\end{algorithm}

\begin{table*}[h]
\centering
\begin{tabular}{|c|c|c|c|}
\hline
PA scheme & Per-step complexity & Total complexity & overhead per-BS (BS j) \\
\hline
Proposed multi-cell PA & 
\begin{tabular}[c]{@{}c@{}}Step 1 $\mathcal{O}(L^3K^2M^2)$\\ Step 2 $\mathcal{O}((LK)^2 \log(LK))$\\ Step 3 min:$\mathcal{O}(\tau_p)$, max:$\sum_{i=1}^{(\tau_p-1)} \mathcal{O}(i)$\\ Step 4 $\mathcal{O}(L^2K^2)$\end{tabular} & \begin{tabular}[c]{@{}c@{}} $\mathcal{O}(L^3K^2M^2+$\\ $(LK)^2 (\log(LK)+1))$\\ $\mathcal{O}(L^3K^2M^2)$ for large M \end{tabular} & \begin{tabular}[c]{@{}c@{}} $\frac{LK}{2}(LK-1)$\\ as $L\to \infty$ it is unbounded \end{tabular} \\
\hline
Proposed scalable multi-cell PA & 
\begin{tabular}[c]{@{}c@{}}Step 1 $\mathcal{O}(\sum_{j=1}^L|{\mathcal I}_j|^2M^2)$\\ Step 2 $\mathcal{O}((LK)^2 \log(LK))$\\ Step 3 min:$\mathcal{O}(\tau_p)$, max:$\sum_{i=1}^{(\tau_p-1)} \mathcal{O}(i)$\\ Step 4 $\mathcal{O}(L^2K^2)$\end{tabular} & 
\begin{tabular}[c]{@{}c@{}} $\mathcal{O}(\sum_{j=1}^L|{\mathcal I}_j|^2M^2+$\\ $(LK)^2 (\log(LK)+1))$\\ $\mathcal{O}(\sum_{j=1}^L|{\mathcal I}_j|^2M^2)$ for large M\\ min $\mathcal{O}(LK^2)$ is achieved at $\gamma=1$ \end{tabular} & 
\begin{tabular}[c]{@{}c@{}} $\frac{|{\mathcal I}_j|}{2}(|{\mathcal I}_j|-1)$\\ as $L\to \infty$ it is bounded \end{tabular} \\
\hline
Single-cell PA in \cite{zhu2016weighted} & 
\begin{tabular}[c]{@{}c@{}}\end{tabular} & 
$\mathcal{O}(\tau_p(LK)^3)$ & 
as $L\to \infty$ it is unbounded \\
\hline
\end{tabular}
\caption{Comparison of computational complexity and information exchange overhead}
\label{tab:pa_comparison}
\end{table*}

\subsection{Computational Complexity and Information Exchange Overhead Analysis}

The total computational complexity of the proposed multi-cell PA scheme (Algorithm 1) is evaluated by decomposing it into four main steps: (step 1) trace computation across user pairs, (step 2) aggregation and sorting of trace terms, (step 3, corresponding to the first while-loop in Algorithm 1) initial orthogonal pilot assignment to strongly interfering user pairs, and (step 4, corresponding to the second while-loop in Algorithm 1) pilot reuse assignment to remaining users. Table \ref{tab:pa_comparison} summarizes the complexity order of each step, along with the total complexity of the scheme. The scalable multi-cell PA scheme (Algorithm 2) follows the same four-step structure as the proposed multi-cell PA scheme, with the key difference being that, in (Step 1), each BS computes trace terms only for a finite subset of users, denoted by ${\mathcal I}_j$, rather than for all users across the network. This reduces the per-BS complexity. The other steps remain structurally identical to those of Algorithm \ref{alg:alg1}. The detailed complexity breakdown for each step, along with the total complexity, is provided in Table \ref{tab:pa_comparison}.

\subsection{Our Proposed Single-Cell PA Scheme}
\label{sec:Single-Cell pilot assignment algorithms}
Recall that the single-cell weighted graph-based PA scheme in \cite{zhu2016weighted} was developed for uncorrelated fading channels and matched filter detection. We extend \cite{zhu2016weighted} to the correlated fading model with M-MMSE detection, using it as a baseline to fairly compare with our proposed multi-cell PA schemes and to highlight the benefits of leveraging spatial correlation matrices. To enable this extension, we define a metric, similar to the one in \cite{zhu2016weighted}, based on our SINR approximation in Theorem 3, which quantifies the pilot contamination severity between two pilot-sharing users, user $jk$ and user $li$ where $l \neq j$. The metric is:

\begin{equation}
	\label{Single-Cell pilot assignment algorithms eq5}
	\zeta_{jk,li}=\frac{\delta_{jlik}^2}{\delta_{jk}^2}+\frac{\delta_{ljki}^2}{\delta_{li}^2}
\end{equation}

\section{Pilot and Data Power Control}\label{sec: Power Control}

\subsection{Pilot Power Control}
In this subsection given a PA, we seek the optimal pilot power allocation among all users such that a weighted sum of all the channel estimation error covariance matrices in the network is minimized, subject to a constraint on average transmit power per user $P_{lk}=\frac{E_{lk}}{\tau_p+\tau_u} \forall l,k$ where $E_{lk}$ is the total energy budget for user $lk$ within one coherence block. Let vector ${\hat {\bf p}}=[{\hat p}_{11},\cdots,{\hat p}_{LK}]$ contain the pilot powers of all users. This constrained optimization problem can be written as the following, where $\breve\omega_{lk}$ is the weight for user $lk$:
\begin{equation}
	\begin{aligned}
		\label{Pilot power control: eq2}
		(P1)\quad\max_{\bf \hat{p}} \quad & {\sum_{(j,l,k)}{\breve\omega_{lk}tr\left({\hat p}_{lk}\mathbf{R}_{lk}^j\left(\boldsymbol{\psi}_{lk}^j\right)^{-1}\mathbf{R}_{lk}^j\right)}}\\
		\textrm{s.t.} \quad & { \hat{p}_{lk}}\le P_{lk}, \forall l,k
	\end{aligned}
\end{equation}
To solve (P1) we utilize the matrix partial fraction transform, and in particular, matrix quadratic transform in \cite{shen2019optimization,shen2024accelerating}, to transform (P1) to an equivalent problem (P2).
\begin{thm}[\cite{shen2019optimization}]
	 \label{Matrix Quadratic Transform}		
		Given a nonempty constraint set X, a sequence of non-negative functions ${\bf A}_n({\bf x})\in\mathbb{H}_{+}^{M\times M}$, strictly positive $\mathbf{B}_n({\bf x})\in \mathbb{H}_{++}^{M\times M}$, and nondecreasing matrix functions $F_n:\mathbb{H}_{+}^{M\times M}\rightarrow\mathbb{R}$, in the sense that $F_n({\bf Z})\geq F_n({\bf Z}^\prime)$ if ${\bf Z}\succeq{\bf Z}^\prime$, for $n=1,\cdots,N$, the sum-of-functions-of-matrix-ratio problem
		\begin{equation}
			\begin{aligned}
				\label{Pilot power control: eq3}
				\max_ {\bf x} \quad & {\sum_{n=1}^{N}{F_n\left({\bf A}_n^H({\bf x}){\bf B}_n^{-1}({\bf x}){\bf A}_n({\bf x})\right)}}\\
				\textrm{s.t.} \quad & { {\bf x}\in X}
			\end{aligned}
		\end{equation}
		is equivalent to
		\begin{equation}
			\begin{aligned}
				\label{Pilot power control: eq4}
				\max_{{\bf x},{\bf\Lambda}_n, \forall n} \quad & {\sum_{n=1}^{N}{F_n\left(2\Re\{{\bf A}_n^H({\bf x})\mathbf{\Lambda}_n\}-\mathbf{\Lambda}_n^H{\bf B}_n({\bf x})\mathbf{\Lambda}_n\right)}}\\
				\textrm{s.t.} \quad & { {\bf x}\in X}, \quad {\mathbf \Lambda_n}\in\mathbb{C}^{M\times M},\forall n
			\end{aligned}
		\end{equation}
with an auxiliary variable $\mathbf{\Lambda}_n$ introduced for each matrix ratio term ${\bf A}_n^H({\bf x}){\bf B}_n^{-1}({\bf x}){\bf A}_n({\bf x})$.
\end{thm}

	Leveraging Theorem \ref{Matrix Quadratic Transform}, we transform (P1) into (P2):
	\begin{equation}
		\begin{aligned}
			\label{Pilot power control: eq5}
			(P2)\quad\max_{{\hat{\bf p}},{\bf\Lambda}_{lk}^j, \forall j,l,k} \quad & {\sum_{(j,l,k)}{tr(2\Re\{\sqrt{{\hat{p}}_{lk}}\mathbf{R}_{lk}^j\mathbf{\Lambda}_{lk}^j\}-{{\bf \Lambda}_{lk}^j}^H\boldsymbol{\psi}_{lk}^j\mathbf{\Lambda}_{lk}^j)}}\\
			\textrm{s.t.} \quad & { {\hat p}_{lk}\le P_{lk} },\forall l,k\quad {\mathbf \Lambda}_{lk}^j\in\mathbb{C}^{M\times M},\forall l,j,k
		\end{aligned}
	\end{equation}
To solve (P2) we decompose it into two sub-problems and we iterate between solving these two sub-problems until we converge to the solution. In the first sub-problem, given $\bf\hat{p}$, we minimize the objective function in (\ref{Pilot power control: eq5}) with respect to ${\mathbf{\Lambda}_{lk}^j}, \forall j,l,k$. Let ${\mathbf{\Lambda}_{lk}^j}^\ast$ be the minimizer of the objevtive function. We obtain:
	\begin{equation}		
		\label{Pilot power control: eq7}
		{\mathbf{\Lambda}_{lk}^j}^\ast=\sqrt{{\hat{p}}_{lk}}(\boldsymbol{\psi}_{lk}^j)^{-1}(\mathbf{R}_{lk}^j)^H
	\end{equation}
In the second sub-problem, given ${\mathbf{\Lambda}_{lk}^j}, \forall l,j,k$, we minimize the objective function in (\ref{Pilot power control: eq5}) with respect to $\hat{p}_{lk}, \forall l,k$. Let $\hat{p}_{lk}^*$ be the minimizer of the objective function. We obtain:
	\begin{equation}		
		\label{Pilot power control: eq8}
		{\hat{p}}_{lk}^\ast=min\{(\frac{\sum_{j} tr(\Re\{\mathbf{R}_{lk}^j\mathbf{\Lambda}_{lk}^j\})}{\tau_\rho\sum_{j}\sum_{(r,i)\in\acute\pi_{lk}}{tr((\mathbf{\Lambda}_{ri}^j)^H\mathbf{R}_{lk}^j\mathbf{\Lambda}_{ri}^j)+\eta_{lk}}})^2,P_{lk}\}
	\end{equation}
	where the Lagrange multiplier $\eta_{lk}$ corresponding to the power constraint is:
	 \begin{equation}\label{Pilot power control: eq9}
	 	{\eta_{lk}^\ast} =
	 	\begin{cases}
	 		0,& if\ {\hat{p}}_{lk}^\ast\le P_{lk},\\
	 		{\eta_{lk}>0,}&{\text{with}\quad {\hat{p}}_{lk}^\ast=P_{lk}}.
	 	\end{cases}
	 \end{equation}

\subsection{Uplink Data Power Control To Maximize Uplink Sum SE}
Given a PA and pilot power allocation, we find the optimal uplink data power allocation among all users such that the weighted sum of all users' rates  is maximized, subject to a constraint on average energy per user in uplink. Let vector ${\bf p}^{ul}=[p_{11}^{ul},\cdots,p_{LK}^{ul}]$ contain the uplink data powers of all users. This constrained optimization problem can be expressed as the following, where $\breve\omega_{jk}$ is the weight for user $jk$:
\begin{equation}
	\begin{aligned}
		\label{Uplink Payload power control: eq1}
		(P3)\quad\max_{{\bf p}^{ul}} \quad & \sum_{(j,k)}\breve\omega_{jk}\overline{SE}_{jk}^{ul}\\
		\textrm{s.t.} \quad & {\hat{p}}_{jk}\tau_p+{p}_{jk}^{ul}\tau_u\le E_{jk}, \forall j,k
	\end{aligned}
\end{equation}
One can show that (P3) is a non-convex and NP-hard. To solve (P3), we apply high SNR approximation to approximate  $log(1+ SINR)$ with $log(SINR)$, which allows us to convert (P3) into a geometric programing (GP) problem. To be specific, let ${\bf q}=[q_{11},\cdots,q_{LK}]^T$ be a vector of auxiliary variables, where $q_{jk} 
\leq SINR_{jk}^{\breve\omega_{jk}}$. We convert (P3) into (P4):
\begin{equation}
	\begin{aligned}
		\label{Uplink Payload power control: eq2}
		(P4)\quad\max_{{\bf p}^{ul}, {\bf q}} \quad & \prod_{(j,k)}{q_{jk}}\\
		\textrm{s.t.} \quad & \frac{{q_{jk}}^\frac{1}{\breve\omega_{jk}}}{p_{jk}{\bf U}_{uu}}\left(\sum_{(l,m)}p_{lm}{\bf F}_{uv}+\sigma_{ul}^2\right)\le1,\\
		&{\hat{p}}_{jk}\tau_p+{p}_{jk}^{ul}\tau_u\le E_{jk}, \forall j,k
	\end{aligned}
\end{equation}
where ${\bf F}_{uv}$ and ${\bf U}_{uu}$ in  (\ref{Uplink Payload power control: eq2}) are the (u,v)-th element and (u,u)-th element of matrix $\mathbf{F}\in\mathbb{R}^{LK\times L K}$ and diagonal matrix $\mathbf{U}\in\mathbb{R}^{LK\times L K}$ given below, and $u=k+(j-1)K$ and $v=m+(l-1)K$. 

\begin{equation}\label{Uplink Payload power control: eq4}
	{\mathbf{F}_{uv}} =
	\begin{cases}
		\,\, 0 \quad,& \text{if}\ (l,m)=(j,k),\\
		{\frac{\delta_{jlmk}^2}{\delta_{jk}^{\prime\prime}}},&{\text{if}\ i_{lm}=i_{jk},(l,m)\neq(j,k)},\\
		{\frac{\mu_{jlmk}}{\delta_{jk}^{\prime\prime}},}&{\text{if}\ i_{lm}\neq i_{jk}},
	\end{cases}
	\quad\mathbf{U}_{uu}=\frac{\delta_{jk}^2}{\delta_{jk}^{\prime\prime}},
\end{equation}
Note that (P4) can solved efficiently by CVX toolbox in MATLAB. For fixed ${\bf F}$ and ${\bf U}$, \cite{tan2012fast} proposed a low complexity  fixed point iteration method, based on KKT optimality condition, that converges geometrically fast to the optimal solution, when the power  coefficient $p_{jk}^{ul}$ is updated as the following (with our notations):
\begin{equation}
		\label{Uplink Payload power control: eq3}
		p_{jk}(t+1)=min\{\sfrac{\breve\omega_{jk}}{(\sum_{(l,m)}\frac{\breve\omega_{lm}\mathbf{F}_{vu}{\overline {SINR}}_{lm}^{ul}({\bf p}^{ul}(t))}{\mathbf{U}_{vv}p_{lm}(t)}),p_{jk}^{ul}(0)}\}
\end{equation}
where $t$ is the iteration index in the fixed point algorithm, for $t=0,1,\cdots$, and the initial point vector is ${\bf p}^{ul}(0)=[p_{11}^{ul}(0),\cdots,p_{LK}^{ul}(0)]^T$ with $p_{jk}^{ul}(0)=(E_{jk}-{\hat p}_{jk}\tau_p)/\tau_u$. Note that in our case $\mathbf{F}$ and $\mathbf{U}$ are not fixed since $\delta_{jk}$, $\mu_{jlmk}$, and $\delta_{jk}^{\prime\prime}$ will change as $p_{jk}^{ul}$ changes. To address this issue  we calculate ${\bf F}$ and ${\bf U}$ only once using the initial point vector ${\bf p}^{ul}(0)$, and keep them fixed during the iteration, and use (\ref{Uplink Payload power control: eq3}) to update $p_{jk}^{ul}\forall j,k$, until the algorithm converges. 
\subsection{Uplink Data Power Control To Maximize the Lowest SE}
Here, we maximize the lowest user rate, to ensure fairness among users. In particular, we consider the following problem:
\begin{equation}
	\begin{aligned}
		\label{Uplink Payload power control: eq6}
		(P5)\quad\max_{{\bf p}^{ul}}\quad\min_{j,k}&  \quad {{\overline {SINR}}_{jk}^{ul}}\\
		\textrm{s.t.} \quad &{\hat{p}}_{jk}\tau_p+{p}_{jk}^{ul}\tau_u\le E_{jk}, \forall j,k
	\end{aligned}
\end{equation}
which can be converted into the following problem:
\begin{equation}
	\begin{aligned}
		\label{Uplink Payload power control: eq6_2}
		(P6)\quad\max_{{\bf p}^{ul}, q}&  \quad q\\
		\textrm{s.t.} \quad & \frac{{q}}{{\overline {SINR}}_{jk}^{ul}}\le1, \quad
		{\hat{p}}_{jk}\tau_p+{p}_{jk}^{ul}\tau_u\le E_{jk}
	\end{aligned}
\end{equation}
where $q$ is an auxiliary variable. An effective strategy to solve (P6) is to apply the bisection method on the parameter $q$. Through the iterative resolution of a succession of linear feasibility problems, this method facilitates rapid convergence towards the global optimal solution of (P6)\cite{bjornson2017massive}.
\subsection{Uplink-Downlink Duality Power Control}
Utilizing the deterministic SINR approximation in (\ref{Asymptotic_Analysis_eq9}), the following theorem establishes a duality relationship between the uplink and downlink scenarios.
\begin{thm}[extension of Theorem 3 in \cite{Bjornson2016MassiveMF}] \label{Theorem: U/D Duality Power Control}
Let vector ${\bf p}^{dl}=[p_{11}^{dl},\cdots,p_{LK}^{dl}]$ include the downlink data powers of all users. Given any set of detectors and uplink data power vector ${\bf p}^{ul}$, $SINR_{jk}^{ul}=SINR_{jk}^{dl}$ and $SE_{jk}^{ul}=SE_{jk}^{dl}, \forall j,k$ if precoders are chosen according to (\ref{UL/DL duality for Deterministic Downlink Rate with M-MMSE eq1}), and downlink data powers are selected as below 
\cite{Bjornson2016MassiveMF}:

	\begin{equation}
		\label{U/D Duality Power Control: eq2}
		{\bf p}^{dl}=\frac{\sigma_{dl}^2}{\sigma_{ul}^2}\left(\mathbf{U}-\mathbf{\Gamma B}^T\right)^{-1}\left(\mathbf{U}-\mathbf{\Gamma B}\right){\bf p}^{ul}
	\end{equation}
	
	where ${\bf\Gamma}=diag({\overline {SINR}}_{11},\ldots,{\overline {SINR}}_{LK})\in\mathbb{C}^{LK\times LK}\in\mathbb{C}^{LK\times LK}$, $\mathbf{B}=\mathbf{F}+\mathbf{U}$, and ${\bf F}$ and ${\bf U}$ are given in (\ref{Uplink Payload power control: eq4}). The proof follows similar steps to the duality proof presented in \cite{Bjornson2016MassiveMF} and is therefore omitted for brevity.
\end{thm}
 \begin{figure*}
	\centering
	\subfloat[\footnotesize{ASD=2, K=10}]{\includegraphics[width=0.68\columnwidth]{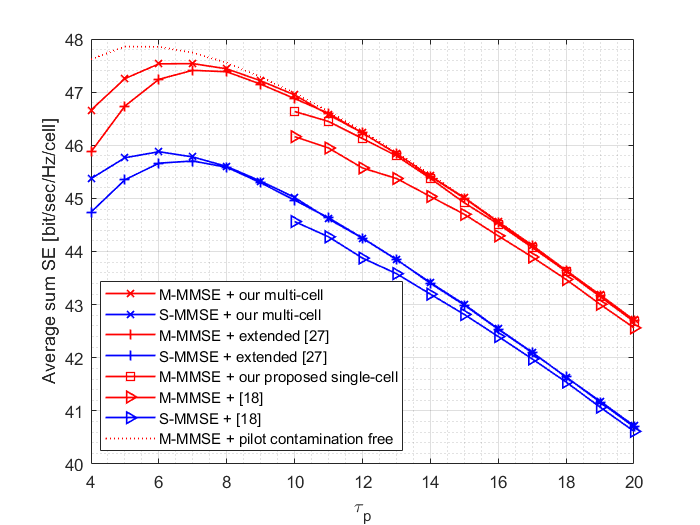}}
	\hfil
	\subfloat[\footnotesize{ASD=10, K=10}]{\includegraphics[width=0.68\columnwidth]{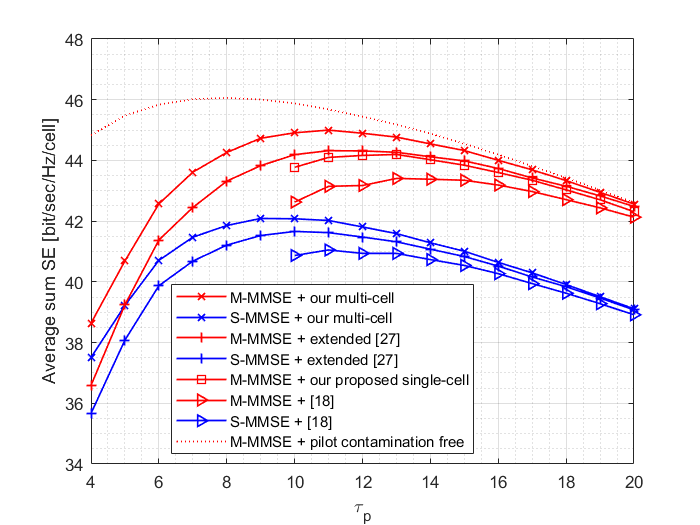}}
	\hfil
	\subfloat[\footnotesize{ASD=20, K=10}]{\includegraphics[width=0.68\columnwidth]{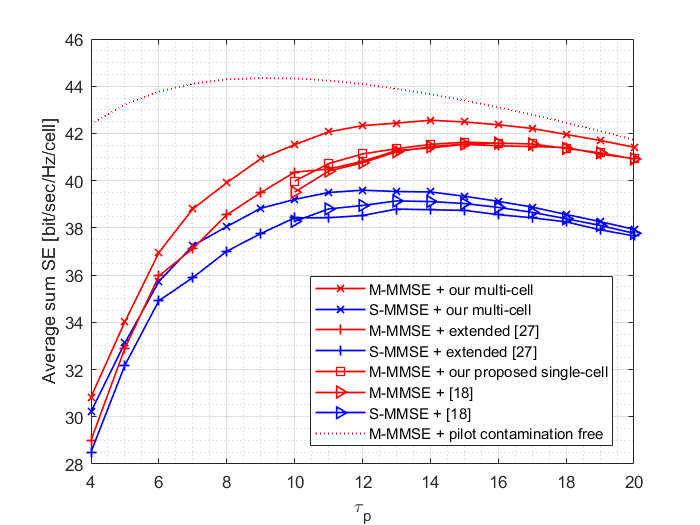}}
	
	\subfloat[\footnotesize{ASD=10, K=30}]{\includegraphics[width=0.68\columnwidth]{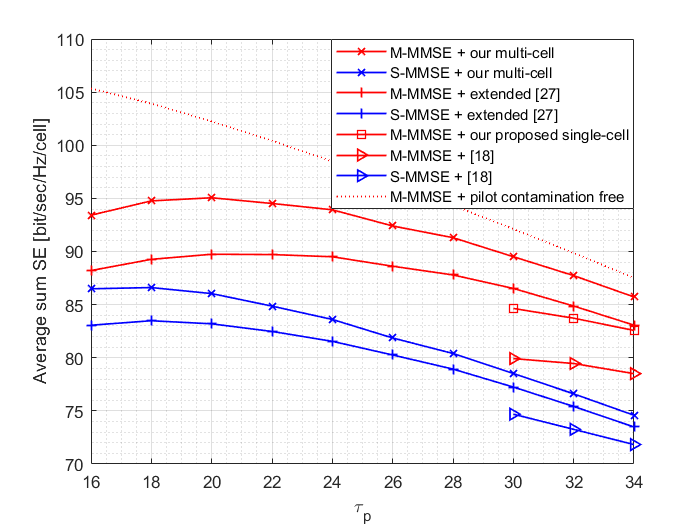}}
	\hfil
	\subfloat[\footnotesize{ASD=20, K=30}]{\includegraphics[width=0.68\columnwidth]{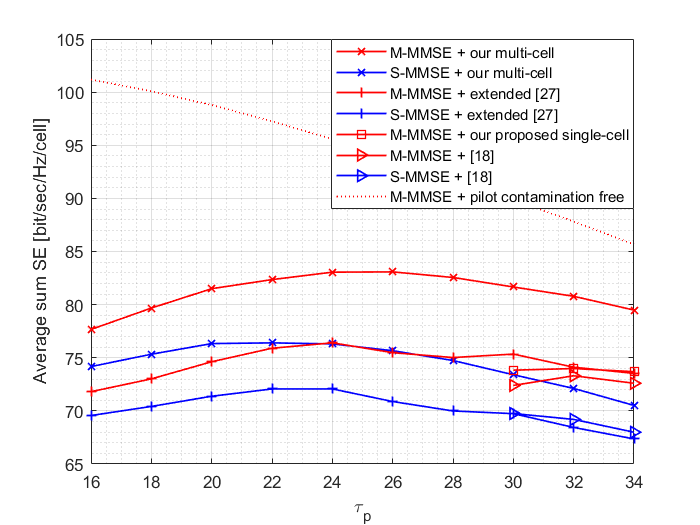}}
	\hfil
	\subfloat[\footnotesize{$r$ = 0.5, K=10}]{\includegraphics[width=0.68\columnwidth]{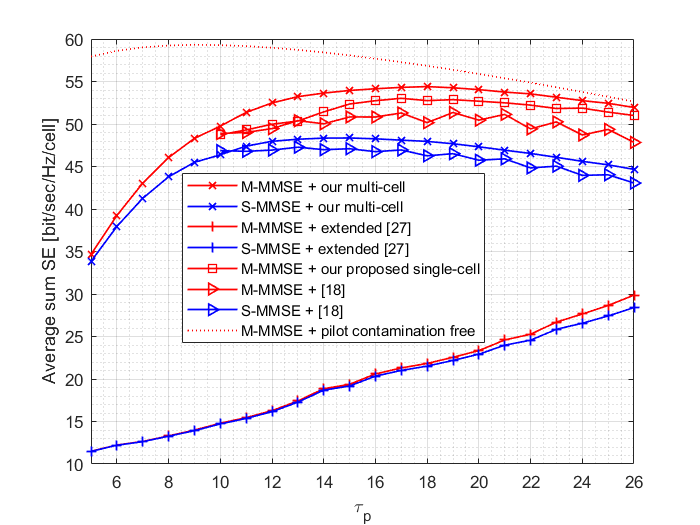}}
	\caption{\footnotesize{Average uplink sum SEs per cell versus $\tau_p$, (a), (b), (c), (d), (e) for Gaussian and (f) for exponential local scattering models, respectively.}}
	\label{fig:n1}
\end{figure*}

\section{Simulation results and discussion}\label{sec: Simulation results and discussion}
In this section, we present simulation results to demonstrate the effectiveness of our proposed PA and power allocation schemes. We consider  $L=4$ square cells, each covering a square of $0.5 km \times 0.5 km$, on a square grid of $2 \times 2$ cells, and simulate a wrap-around topology to mitigate edge effects and ensure uniform simulation performance across all cells\cite{bjornson2017massive}. A BS with $M=100$ antennas is located at the center of each cell. The users are uniformly and independently distributed within each cell, positioned at distances greater than $35m$ from the serving BS. We let the system bandwidth $W=20MHz$ and, $\tau_c=200$ transmission symbols, and $\tau_d=\tau_u+\tau_p=\frac{\tau_c}{2}$. Due to uplink-downlink Duality we only consider uplink transmission for brevity. We assume the receiver noise power is $-94 dBm$, $E_{lk}=20J$, and per-user transmit power is $P=20dBm$ (except Fig. 2 and 3). With these parameters and equal power allocation among users, the received SNR at the BS from users positioned at the vertex of a cell is approximately $13dB$, accounting for path loss only. We adopt 3GPP LTE model in $2GHz$ carriers for $\beta_{lk}^j$ (in dB) \cite{LTE3GPP}:
\begin{equation}
	\label{Simulation results and discussion: eq1}
	\beta_{lk}^j(in\, dB)=-148.1-37.6\log_{10}{(\frac{d_{lk}^j}{1km})}+z_{lk}^j
\end{equation}
where $d_{lk}^j$ in $km$ measures the distance between user $lk$ and BS $j$, and $z_{lk}^j\sim\mathcal{N}(0,10dB)$ models the log-normal shadowing effect. To characterize spatial correlation matrix ${\bf R}$ for ULA antennas in terms of the large scale fading coefficient $\beta$ we consider the following two models:

\quad$\bullet$ For Gaussian  local scattering model, the $(n,m)$-th entry of ${\bf R}$ \cite{bjornson2017massive} is $[\mathbf{R}]_{n,m}=\beta\int{e}^{2\pi j d_H(n-m)\sin{({\phi})}}f(\phi)d\phi$. The angle  of a multipath component is $\phi= \bar\phi+ \delta$, where $\bar\phi$ is a deterministic nominal angle, $\delta\sim \mathcal{N}(0,\sigma_{\bar\phi}^2)$, and $\sigma_{\bar\phi}$ (measured in radians) is the angular standard deviation (ASD).

\quad$\bullet$ For exponential  local scattering model, the $(n,m)$-th  entry of ${\bf R}$ is $[\mathbf{R}]_{n,m}=\beta \textit{r}^{|n-m|} e^{i(n-m)\bar\phi}$ \cite{loyka2001channel}, where $r\in[0,1]$ is called the correlation factor.

To demonstrate the benefits of our proposed multi-cell PA schemes, we compare them with the single-cell PA schemes in \cite{zhu2016weighted}, our proposed single-cell PA scheme, extension of the single-cell PA scheme in \cite{you2015pilot} with condition (\ref{Multi-Cell pilot assignment algorithms eq14}) to a multi-cell PA scheme using Algorithm 1, (dubbed as extended \cite{you2015pilot} scheme), with S-MMSE, P-MMSE, and M-MMSE processors, and use pilot contamination free for M-MMSE as a benchmark.

 \begin{figure*} 
	\centering
	\subfloat[\footnotesize{ASD = 2, $\tau_p = K$ }]{\includegraphics[width=0.68\columnwidth]{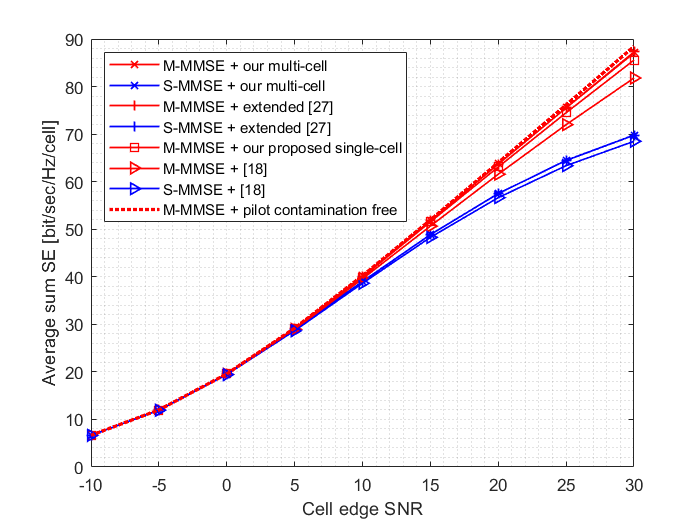}}
	\hfil
	\subfloat[\footnotesize{ASD = 20, $\tau_p = K$}]{\includegraphics[width=0.68\columnwidth]{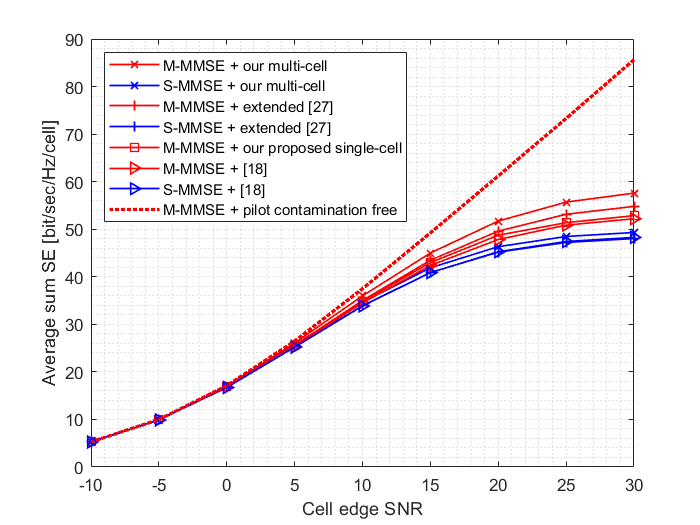}}
	\hfil
	\subfloat[\footnotesize{ASD = 20, $\tau_p = 2K$}]{\includegraphics[width=0.68\columnwidth]{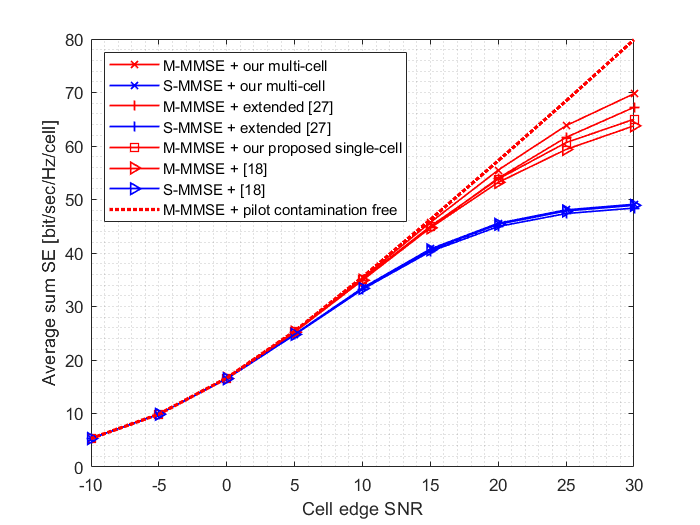}}
	\caption{\footnotesize{Average uplink sum SE per cell versus cell edge SINR for Gaussian scattering model and $K=10$.}}
	\label{fig:n4}
\end{figure*}


\begin{figure*}[t]
    \centering
    \setcounter{subfigure}{0}
    \subfloat[ASD = 5]{%
        \includegraphics[width=0.48\linewidth]{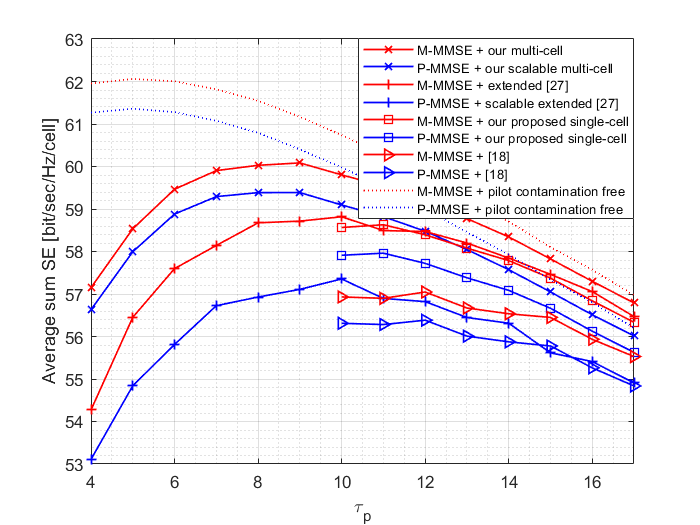}
    }
    \hspace{-5mm}
    \subfloat[ASD = 10]{%
        \includegraphics[width=0.48\linewidth]{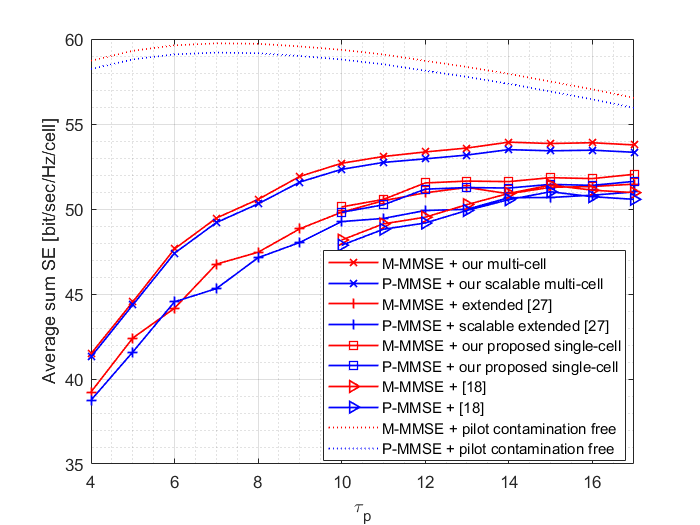}
    }
    
    \caption{\scriptsize Average uplink sum SE per cell versus $\tau_p$ for Gaussian scattering model with\\ $K=10$, $L=16$, each covering a square of $0.5\,\text{km}\times0.5\,\text{km}$, and $P=24.8\,\text{dBm}$.}
    \label{fig:taup}
\end{figure*}

\begin{figure*}[t]
    \centering
    \setcounter{subfigure}{0}
    \subfloat[ASD = 10]{%
        \includegraphics[width=0.55\linewidth]{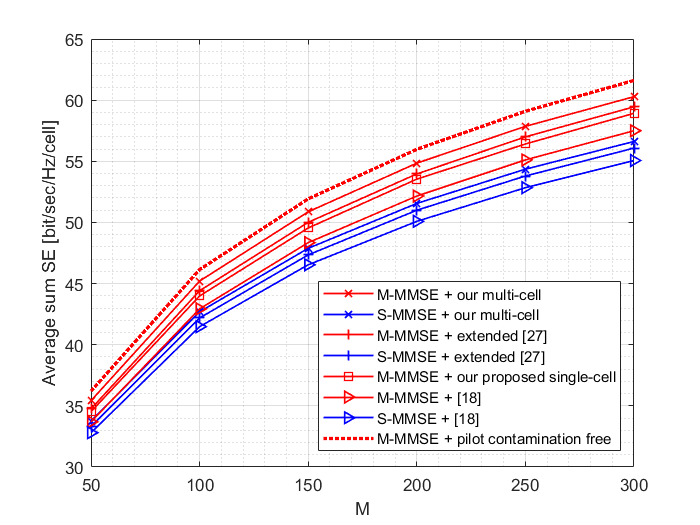}
    }
    \hspace{-5mm}
    \subfloat[ASD = 10]{%
        \includegraphics[width=0.20\linewidth]{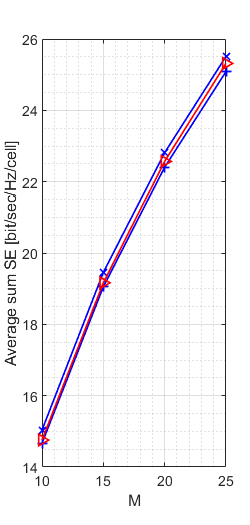}
    }
    \hspace{-3mm}
    \subfloat[ASD = 20]{%
        \includegraphics[width=0.20\linewidth]{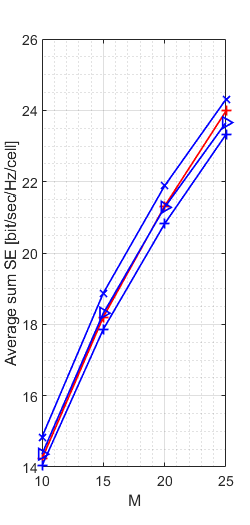}
    }

    \caption{\footnotesize Average uplink sum SE per cell versus $M$ for Gaussian scattering model with $ASD=10$ and $\tau_p = K = 10$.}
    \label{fig:Mplots}
\end{figure*}

\setcounter{figure}{4}
\begin{figure}[h]  
    \centering
    \includegraphics[width=0.50\textwidth]{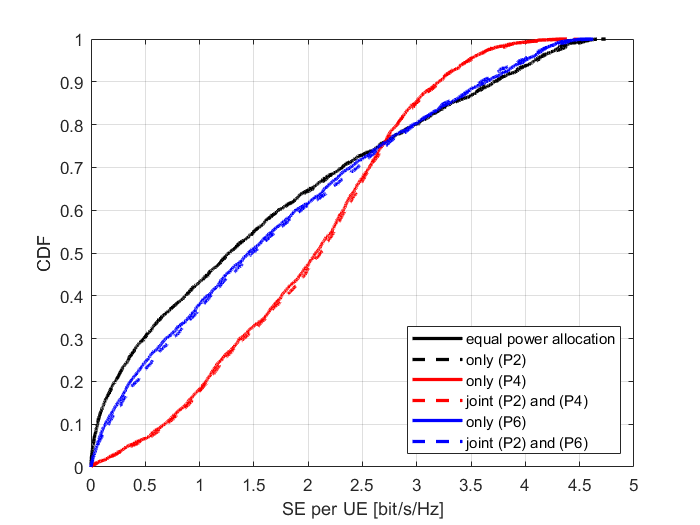}  
    \caption{\footnotesize{CDF of per-user SE for Gaussian scattering model, $ASD=10$ and $\tau_p = K=10$.}}
    \label{fig:nm6}
\end{figure}
\addtocounter{figure}{1}

Fig. (\ref{fig:n1}) plots uplink sum SE versus $\tau_p$ for Gaussian and exponential local scattering models. Fig. 1(a) shows for low ASD different multi-cell PA schemes exhibit similar performance, as there is a reduced likelihood of overlapping multipath AOA intervals among users sharing pilot, resulting in minimal coherent interference and mitigating the impact of PA schemes. However, as the $\tau_p$ decreases, the performance of different schemes becomes increasingly distinct and divergent. Figs. 1(b) and 1(c) show as ASD increases (resulting in heightened coherent interference), the superiority of our proposed multi-cell scheme, which directly minimizes this interference,  becomes increasingly evident. We observe that the optimal $\tau_p$ value, at which the sum SE is maximized, increases with ASD. However, the maximized sum SE decreases accordingly. Also, our single-cell scheme consistently outperforms the scheme in \cite{zhu2016weighted}. However, as ASD increases, and the correlated channel model approaches the uncorrelated channel model, their performance begins to converge closely. Figs. 1(d) and 1(e) show as  $K$ increases, our multi-cell PA scheme exhibits a remarkable performance superiority compared to alternative approaches. Noticeably, the fusion of S-MMSE and our multi-cell scheme surpasses the performance of combining M-MMSE with other PA schemes, underscoring the significance of our orthogonality condition as delineated in (\ref{Multi-Cell pilot assignment algorithms eq4}). Moreover, our multi-cell scheme achieves the maximum sum SE at  a considerably lower $\tau_p$ compared to $K$. Fig. \ref{fig:n1} (f) plots uplink sum SE versus $\tau_p$ for Exponential local scattering model, which generates full-rank spatial correlation matrices. Our proposed multi-cell PA scheme significantly outperforms the extended \cite{you2015pilot}, demonstrating its robustness even under full-rank spatial correlation conditions. In the remaining, we focus on Gaussian local scattering model.

Fig. (\ref{fig:n4}) illustrates the relationship between uplink sum SE and per-user transmit power across various ASD and $\tau_p$ values. With $\tau=K$, as ASD increases, the sum SE begins to plateau at high transmit power levels. To counteract this saturation effect, increasing $\tau_p$ can be effective, boosting the sum SE. The advantages of utilizing a multi-cell PA strategy become evident when contrasting the integration of a multi-cell scheme with M-MMSE and S-MMSE processing, as depicted in Fig. (\ref{fig:n4}b) and Fig. (\ref{fig:n4}c), particularly at high transmit power levels. While the sum SE increase remains limited with S-MMSE processing, there is a notable escalation in sum SE with M-MMSE processing as transmit power rises.

Fig. (4) plots uplink sum SE versus $M$. As $M$ increases, the potential for achieving a larger sum SE with M-MMSE processing compared to S-MMSE processing becomes more pronounced. Surprisingly, with small $M$, the combination of S-MMSE with our multi-cell PA scheme matches or even outperforms the integration of M-MMSE with the scheme in \cite{zhu2016weighted}. This is because the scheme in \cite{zhu2016weighted} relies on asymptotic SINR approximations under i.i.d. Rayleigh fading and large $M$ assumptions, which break down when M is small. In contrast, our multi-cell PA scheme exploits spatial correlation matrices, remaining effective even with small $M$ and high angular spreads where orthogonal subspaces are scarce. Focusing on M-MMSE processing, as $M$ increases, the sum SE of our single-cell scheme experiences more growth compared to the scheme in \cite{zhu2016weighted}, while the sum SE of the extended \cite{you2015pilot} scheme shows even more growth compared to our single-cell scheme. Furthermore, our multi-cell scheme demonstrates even higher sum SE growth compared to all of the aforementioned techniques. To evaluate the scalability of the proposed scalable multi-cell PA scheme, we consider a larger network with 16 cells arranged in a $4\times4$ grid ($2km \times 2km$ total area), introducing greater variability in large-scale fading. By setting a tunable threshold $\gamma = 0.013$, our scalable PA scheme adaptively limits the subset of users included in processing at each BS. Fig. (3) confirms that even with P-MMSE processing, the scalable PA scheme achieves near-optimal SE performance, closely matching that of the full M-MMSE-based multi-cell PA scheme across different $ASD$ and $\tau_p$ values. In contrast, scalable extension of PA scheme in \cite{you2015pilot} suffers from notable performance degradation, particularly under high angular spreads where spatial correlation matrices approach full rank.

\begin{figure*}[b]
   \centering
    \rule{\textwidth}{0.4pt}
   \begin{equation}
   \label{lemma4-1}
    {\bf F}_{n}^{\pi_{jk}}({\bf A},{\bf X},{\bf Y},{\bf B})={{\bf F}_{n-1}^{\pi_{jk}}({\bf A},{\bf X},{\bf Y},{\bf B})-\frac{p_{mr}{\bf F}_{n-1}^{\pi_{jk}}({\bf A},{\bf X},\mathbf{\Sigma},{\bf I}){\hat{{\bf h}}}_{mr}^j({\hat{{\bf h}}}_{mr}^j)^H{\bf F}_{n-1}^{\pi_{jk}}({\bf I},\mathbf{\Sigma},{\bf Y},{\bf B})}{1+p_{mr}({\hat{{\bf h}}}_{mr}^j)^H{\bf F}_{n-1}^{\pi_{jk}}({\bf I},\mathbf{\Sigma},\mathbf{\Sigma},{\bf I}){\hat{{\bf h}}}_{mr}^j}} 
    \end{equation}
    \rule{\textwidth}{0.4pt}
    \begin{equation}
    \label{lemma4-2}
            {\bf {\acute F}}_{n}^{\pi_{jk}}({\bf A},{\bf X},{\bf Y},{\bf B})={{\bf {\acute F}}_{n-1}^{\pi_{jk}}({\bf A},{\bf X},{\bf Y},{\bf B})}-\frac{{p_{mr}}{\bf {\acute F}}_{n-1}^{\pi_{jk}}({\bf A},{\bf X},\mathbf{\Sigma},{{\bf \Upsilon}_{mr}^{jk}}){\hat{\bf h}}_{jk}^j({\hat{\bf h}}_{jk}^j)^H{\bf {\acute F}}_{n-1}^{\pi_{jk}}({{\bf \Upsilon}_{mr}^{jk}}^H,\mathbf{\Sigma},{\bf Y},{\bf B})}{1+p_{mr}({\hat{\bf h}}_{jk}^j)^H{\bf {\acute F}}_{n-1}^{\pi_{jk}}({{\bf \Upsilon}_{mr}^{jk}}^H,\mathbf{\Sigma},\mathbf{\Sigma},{\bf \Upsilon}_{mr}^{jk}){\hat{\bf h}}_{jk}^j}    
    \end{equation}
    \rule{\textwidth}{0.4pt}
    \begin{equation}
    \label{lemma4-3}
            {{\tilde f}}_{n}^{\pi_{jk}}({\bf\Xi}_{jk,jk}^j,{\bf A},{\bf X}, {\bf Y}, {\bf B})={{\tilde f}}_{n-1}^{\pi_{jk}}({\bf\Xi}_{jk,jk}^j,{\bf A},{\bf X}, {\bf Y}, {\bf B})-\frac{p_{mr}{{\tilde f}}_{n-1}^{\pi_{jk}}({\bf\Xi}_{jk,jk}^j,{\bf A},{\bf X},\mathbf{\Sigma}, {\bf \Upsilon}_{mr}^{jk}){{\tilde f}}_{n-1}^{\pi_{jk}}({\bf\Xi}_{jk,jk}^j,{{\bf\Upsilon}_{mr}^{jk}}^H,\mathbf{\Sigma}, {\bf Y}, {\bf B})}{1+{\hat{p}}_{mr}{ {\tilde f}}_{n-1}^{\pi_{jk}}({\bf\Xi}_{jk,jk}^j,{{\bf \Upsilon}_{mr}^{jk}}^H,\mathbf{\Sigma},\mathbf{\Sigma}, {\bf \Upsilon}_{mr}^{jk})}
    \end{equation}
    \end{figure*}


To investigate the effectiveness of the proposed pilot and data power allocation, Fig.(\ref{fig:nm6}) displays the cumulative distribution function (CDF) of per-user SE. As a baseline, we provide the results for equal power allocation where $p_{lk}=\hat{p}_{lk}=P$. By solely optimizing data power allocation through solving (P6), fairness among users notable improve compared to the baseline. When optimizing only data power allocation by solving (P4) results in a significant increase in sum SE, 20.3\%, and fairness compared to the baseline. When optimizing both pilot and data power allocation by solving (P2) and (P4) or (P2) and (P6), there is only a slight improvement in the resulting sum SE compared to optimizing only data power without pilot power optimization (by solving only (P4) or only (P6)).

\section{CONCLUSION}\label{CONCLUSION}
In this work, we investigated a multi-cell massive MIMO system with spatially correlated Rayleigh fading, pilot reuse across the network, and M-MMSE processing at the BSs. We derived a novel deterministic approximation of the uplink SINR (Theorem 3), valid under pilot reuse and spatial correlation, addressing a key gap in the literature where such closed-form approximations were previously unavailable, even for cell-free systems. Building on this result, we developed a new multi-cell PA scheme that fully eliminates pilot contamination by exploiting the spatial correlation matrices of all users. To ensure scalability in large networks, we introduced a scalable extension with P-MMSE processing, which significantly reduces inter-BS information exchange while maintaining near-optimal SE. We also designed joint pilot and data power allocation schemes under both weighted sum SE and max-min SE objectives, supported by a detailed complexity analysis confirming their practicality. Simulation results offer system design guidelines: for moderate network sizes, the multi-cell PA with M-MMSE achieves the highest SE by eliminating pilot contamination, while the scalable PA with P-MMSE is preferable in larger networks due to its lower information exchange overhead. Our results also show that large pilot reuse factors, as in \cite{li2017massive}, are impractical under typical outdoor deployments with mobility and realistic
coherence block sizes (e.g., $\tau_c = 200$ symbols), while our PA schemes enable effective reuse with minimal overhead.
\appendices
\section{USEFUL LEMMAS}
\label{Appendix:app A}
\begin{lem}[Lem. B.26 \cite{bai2010spectral}, Thm. 3.7 \cite{couillet2011random}, Lem. 12 \cite{hoydis2012random}]\label{Lemma: Appendix A lem1}
	Let ${\bf A}\in\mathbb{C}^{M\times M}$ and $x,y\sim \mathcal{N_C}(0,\frac{1}{M}I_M)$. Assume that ${\bf A}$ has a uniformly bounded spectral norm with respect to M and $x$,$y$ are mutually independent and independent of ${\bf A}$. Then:
	\begin{enumerate}
		\item{$\{|(x^H{\bf A}x)^2-(\frac{1}{M}tr{\bf A})^2|\}\stackrel{M\rightarrow\infty}{\rightarrow}0$}
		\item{$x^H\ {\bf A}\ x-\frac{1}{M}tr{\bf A}\stackrel{\stackrel{\text{a.s.}}{M\rightarrow\infty}}{\longrightarrow}0$}
		\item{$x^H{\bf A}y\stackrel{\stackrel{\text{a.s.}}{M\rightarrow\infty}}{\longrightarrow}0$}
	\end{enumerate}
\end{lem}
\begin{lem}[Matrix inversion Lemma \cite{silverstein1995empirical}]\label{Lemma: Appendix A lem2}
	Let ${\bf A}\in\mathbb{C}^{M\times M}$ a Hermitian invertible matrix. Then, for any vector $x\in\mathbb{C}^M$ and scalar $\tau\in\mathbb{C}$ such that ${\bf A}+\tau\ xx^H$ is invertible, we have:
	\begin{enumerate}
		\item{$	x^H(A+\tau\ xx^H)^{-1}=\frac{x^HA^{-1}}{1+\tau x^HA^{-1}\ x}$}
		\item{$	(A+\tau\ xx^H)^{-1}=A^{-1}-\frac{A^{-1}\tau\ xx^HA^{-1}}{1+\tau x^HA^{-1}\ x}$}
	\end{enumerate}
\end{lem}
\begin{lem}[Rank-1 perturbation Lemma \cite{silverstein1995empirical}]\label{Lemma: Appendix A lem3}
	Let $z<0$, ${\bf A} ,{\bf B}\in\mathbb{C}^{M\times M}$ where ${\bf A}$ is a Hermitian and non-negative definite matrix, and ${\bf v}\in\mathbb{C}^{M}$. Then:
	
	$\mid tr( ({\bf A}-z{\bf I}_M)^{-1}-({\bf A}+{\bf v}{\bf v}^H-z{\bf I}_M)^{-1}{\bf B} ) \mid\leq\frac{\|{\bf B}\|}{\mid z\mid}$
\end{lem}
\begin{lem}\label{Lemma: Appendix A lem4}
	Let ${\bf A}$,${\bf B}$,${\bf X}$, ${\bf Y}$, ${\bf \Upsilon}\in\mathbb{C}^{M\times M}$, $N=|\pi_{jk}|$, and $\pi_{jk_n}$ represent the $n$-th element of set $\pi_{jk}$, and user $mr$ corresponds to index  
	$\pi_{jk_n}$. We define the basic functions:
    $${\bf F}_{0}^{\pi_{jk}}({\bf A},{\bf X}, \mathbf{\Sigma}, {\bf B})={\bf \acute F}_{0}^{\pi_{jk}}({\bf A},{\bf X}, \mathbf{\Sigma}, {\bf B})={\bf AXB},$$ $${\bf F}_{0}^{\pi_{jk}}({\bf A},\mathbf{\Sigma}, {\bf Y}, {\bf B})={\bf \acute F}_{0}^{\pi_{jk}}({\bf A},\mathbf{\Sigma}, {\bf Y}, {\bf B})={\bf AYB},$$  $${{\tilde f}}_{0}^{\pi_{jk}}({\bf\Xi}_{jk,jk}^j,{\bf A},{\bf X}, \mathbf{\Sigma}, {\bf B})=tr({\bf\Xi}_{jk,jk}^j{\bf A}{\bf X}{\bf B}),$$ $${{\tilde f}}_{0}^{\pi_{jk}}({\bf\Xi}_{jk,jk}^j,{\bf A},\mathbf{\Sigma}, {\bf Y}, {\bf B})=tr({\bf\Xi}_{jk,jk}^j{\bf A}{\bf Y}{\bf B}).$$
    The functions ${\bf F}_{n}^{\pi_{jk}}({\bf A},{\bf X}, {\bf Y}, {\bf B})$, ${\bf \acute F}_{n}^{\pi_{jk}}({\bf A},{\bf X}, {\bf Y}, {\bf B})$, and ${{\tilde f}}_{n}^{\pi_{jk}}({\bf\Xi}_{jk,jk}^j,{\bf A},{\bf X}, {\bf Y}, {\bf B})$ satisfy the recursive equations (\ref{lemma4-1}), (\ref{lemma4-2}), and (\ref{lemma4-3}) while $\mathbf{\Sigma}$ is equal to at least one of ${\bf X}$ or ${\bf Y}$. To reduce the size of the equations, we define:
    $${\bf F}_{0}^{\pi_{jk}}({\bf A},{\bf X}, {\bf X}, {\bf B})={\bf F}_{0}^{\pi_{jk}}({\bf A},{\bf X}, {\bf B}),$$
    $${\bf \acute F}_{0}^{\pi_{jk}}({\bf A},{\bf X}, {\bf X}, {\bf B})={\bf \acute F}_{0}^{\pi_{jk}}({\bf A}, {\bf X}, {\bf B}),$$
    $${{\tilde f}}_{0}^{\pi_{jk}}({\bf\Xi}_{jk,jk}^j,{\bf A},{\bf X}, {\bf X}, {\bf B})={{\tilde f}}_{0}^{\pi_{jk}}({\bf\Xi}_{jk,jk}^j,{\bf A},{\bf X}, {\bf B}).$$ Utilizing these recursive formula we can express the functions ${\bf F}_{N}^{\pi_{jk}}$, ${\bf {\acute F}}_{N}^{\pi_{jk}}$, and ${{\tilde f}}_{N}^{\pi_{jk}}$ according to (\ref{APPENDIX B: eqF}), (\ref{APPENDIX B: eq6xxx3}), and (\ref{APPENDIX B: eq6xxx5}), in which all the functions can be expressed in terms of the basic functions ${\bf F}_{0}^{\pi_{jk}}$, ${\bf \acute F}_{0}^{\pi_{jk}}$, and ${{\tilde f}}_{0}^{\pi_{jk}}$. 
\end{lem}
\section{}
\label{Appendix:app B}
\subsection{We define the following matrices for \texorpdfstring{$\forall j,k$}{∀ j,k}}
\begin{enumerate}
	\item{${\boldsymbol{\Sigma}}_{\pi_{jk}}=(\mathbf{\Sigma}_{j}^{-1}-\sum_{n=1,(m,r)=\pi_{jk_n}}^{N} p_{mr}{\hat{\bf{h}}}_{mr}^j({\hat{\bf{h}}}_{mr}^j)^H)^{-1}$}
	\item{${\boldsymbol{\Sigma}}_{\acute\pi_{jk}}=(\mathbf{\Sigma}_{j}^{-1}-\sum_{n=1,(m,r)=\acute\pi_{jk_n}}^{N+1} p_{mr}{\hat{\bf{h}}}_{mr}^j({\hat{\bf{h}}}_{mr}^j)^H)^{-1}$}
	\item$\acute{\boldsymbol{\Sigma}_j}=M{\boldsymbol{\Sigma}}_j$, ${\boldsymbol{\Sigma}}_{jk}=(\mathbf{\Sigma}_{j}^{-1}-p_{jk}{\hat{\bf{h}}}_{jk}^j({\hat{\bf{h}}}_{jk}^j)^H)^{-1}$
\end{enumerate}
\subsection{Power of desired signal in (\ref{Uplink_Data_Transmission_Phase_eq2})}
To calculate the desired signal power, we have:    
\begin{figure*}
   \centering
\begin{equation}
	\label{APPENDIX B: eqF}
	{\bf F}_{N}^{\pi_{jk}}({\bf A},{\bf X},{\bf Y},{\bf B})={{\bf F}_{0}^{\pi_{jk}}({\bf A},{\bf X},{\bf Y},{\bf B})-\sum_{\overset{n=1}{(mr)=\pi_{jk_n}}}^{N}\frac{p_{mr}{\bf F}_{n-1}^{\pi_{jk}}({\bf A},{\bf X},\mathbf{\Sigma},{\bf I}){\hat{{\bf h}}}_{mr}^j({\hat{{\bf h}}}_{mr}^j)^H{\bf F}_{n-1}^{\pi_{jk}}({\bf I},\mathbf{\Sigma},{\bf Y},{\bf B})}{1+p_{mr}({\hat{{\bf h}}}_{mr}^j)^H{\bf F}_{n-1}^{\pi_{jk}}({\bf I},\mathbf{\Sigma},\mathbf{\Sigma},{\bf I}){\hat{{\bf h}}}_{mr}^j}}
\end{equation}

\rule{\textwidth}{0.4pt}

\begin{equation}
	\label{APPENDIX B: eq6xxx3}
	{\bf {\acute F}}_{N}^{\pi_{jk}}({\bf A},{\bf X},{\bf Y},{\bf B})={{\bf {\acute F}}_{0}^{\pi_{jk}}({\bf A},{\bf X},{\bf Y},{\bf B})-\sum_{\overset{n=1}{(mr)=\pi_{jk_n}}}^{N}\frac{{p_{mr}}{\bf {\acute F}}_{n-1}^{\pi_{jk}}({\bf A},{\bf X},\mathbf{\Sigma},{{\bf \Upsilon}_{mr}^{jk}}){\hat{\bf h}}_{jk}^j({\hat{\bf h}}_{jk}^j)^H{\bf {\acute F}}_{n-1}^{\pi_{jk}}({{\bf \Upsilon}_{mr}^{jk}},\mathbf{\Sigma},{\bf Y},{\bf B})}{1+p_{mr}({\hat{\bf h}}_{jk}^j)^H{\bf {\acute F}}_{n-1}^{\pi_{jk}}({\bf \Upsilon}_{mr}^{jk},\mathbf{\Sigma},\mathbf{\Sigma},{\bf \Upsilon}_{mr}^{jk}){\hat{\bf h}}_{jk}^j}}
\end{equation}

\rule{\textwidth}{0.4pt}

\begin{equation}
	\begin{split}
		\label{APPENDIX B: eq6xxx5}
		{{\tilde f}}_{N}^{\pi_{jk}}({\bf\Xi}_{jk,jk}^j,{\bf A},{\bf X},{\bf Y},{\bf B})={{\tilde f}}_{0}^{\pi_{jk}}({\bf\Xi}_{jk,jk}^j,{\bf A},{\bf X},{\bf Y},{\bf B})-\sum_{\overset{n=1}{(mr)=\pi_{jk_n}}}^{N}\frac{p_{mr}{{\tilde f}}_{n-1}^{\pi_{jk}}({\bf\Xi}_{jk,jk}^j,{\bf A},{\bf X},\mathbf{\Sigma},{\bf \Upsilon}_{mr}^{jk}){{\tilde f}}_{n-1}^{\pi_{jk}}({\bf\Xi}_{jk,jk}^j,{{\bf\Upsilon}_{mr}^{jk}}^H,\mathbf{\Sigma},{\bf Y},{\bf B})}{1+{{p}}_{mr}{ {\tilde f}}_{n-1}^{\pi_{jk}}({\bf\Xi}_{jk,jk}^j,{{\bf \Upsilon}_{mr}^{jk}}^H,\mathbf{\Sigma},\mathbf{\Sigma},{\bf \Upsilon}_{mr}^{jk})}
	\end{split}
\end{equation}\rule{\textwidth}{0.2pt}
    \end{figure*}

\begin{equation}
	\begin{split}
		\label{APPENDIX B: eq6xxx1}
		&(\mathbf{v}_{jk})^H{\hat{{\bf h}}}_{jk}^j=(\hat{{\bf h}}_{jk}^j)^H{\boldsymbol{\Sigma}}_j{\hat{{\bf h}}}_{jk}^j \stackrel{\text{(a)}}{=}
		\frac{({\hat{{\bf h}}}_{jk}^j)^H\mathbf{\Sigma}_{jk} {\hat{{\bf h}}}_{jk}^j}{1+p_{jk}({\hat{{\bf h}}}_{jk}^j)^H \mathbf{\Sigma}_{jk}{\hat{{\bf h}}}_{jk}^j}\\
		&\stackrel{\text{(b)}}{=}\frac{({\hat{{\bf h}}}_{jk}^j)^H\mathbf{F}_{N}^{\pi_{jk}}({\bf I},\mathbf{\Sigma}_{\acute\pi_{jk}},{\bf I}) {\hat{{\bf h}}}_{jk}^j}{1+p_{jk}({\hat{{\bf h}}}_{jk}^j)^H \mathbf{F}_{N}^{\pi_{jk}}({\bf I},\mathbf{\Sigma}_{\acute\pi_{jk}},{\bf I}){\hat{{\bf h}}}_{jk}^j}\\
		&\stackrel{\text{(c)}}{=}\frac{{\tilde f}_{N}^{\pi_{jk}}({\bf\Xi}_{jk,jk}^j,{\bf I},\frac{1}{M}\acute{\mathbf{\Sigma}}_{\acute\pi_{jk}},{\bf I})}{1+p_{jk}{\tilde f}_{N}^{\pi_{jk}}({\bf\Xi}_{jk,jk}^j,{\bf I},\frac{1}{M}\acute{\mathbf{\Sigma}}_{\acute\pi_{jk}},{\bf I})}\\
		&\stackrel{\text{(d)}}{=}\frac{{\tilde f}_{N}^{\pi_{jk}}({\bf\Xi}_{jk,jk}^j,{\bf I},\frac{1}{M}\acute{\mathbf{\Sigma}}_{j}{\bf I})}{1+p_{jk}{\tilde f}_{N}^{\pi_{jk}}({\bf\Xi}_{jk,jk}^j,{\bf I},\frac{1}{M}\acute{\mathbf{\Sigma}}_{j}{\bf I})}\\
		&\stackrel{\text{(e)}}{\asymp}\frac{{\tilde f}_{N}^{\pi_{jk}}({\bf\Xi}_{jk,jk}^j,{\bf I},\frac{1}{M}{{\bf T}_j},{\bf I})}{1+p_{jk}{\tilde f}_{N}^{\pi_{jk}}({\bf\Xi}_{jk,jk}^j,{\bf I},\frac{1}{M}{{\bf T}_j},{\bf I})}\stackrel{\text{(f)}}{=}\frac{\delta_{jk}}{1+p_{jk}\delta_{jk}}		
	\end{split}
\end{equation}
where (a) follows from applying Lemma 2.1 to remove $\hat{{\bf h}}_{jk}^j$ from ${\boldsymbol{\Sigma}}_j$, such that ${\boldsymbol{\Sigma}}_{jk}$ and $\hat{{\bf h}}_{jk}^j$ are uncorrelated, (b) follows from using Lemma 2.2 multiple  times to remove all channel estimates corresponding to users who share the same pilot as user $jk$ from ${\boldsymbol{\Sigma}}_{jk}$ and $\mathbf{F}_{N}^{\pi_{jk}}({\bf I},\mathbf{\Sigma}_{\acute\pi_{jk}},{\bf I})$ is obtained using (\ref{APPENDIX B: eqF}), (c) follows from replacing ${\mathbf{\Sigma}_{\acute\pi_{jk}}}$ with $\frac{1}{M}\acute{\mathbf{\Sigma}}_{\acute\pi_{jk}}$ and ${\hat{\bf h}}_{mr}^j={\bf \Upsilon}_{mr}^{jk}{\hat{\bf h}}_{jk}^j$ into $\mathbf{F}_{N}^{\pi_{jk}}({\bf I},\mathbf{\Sigma}_{\acute\pi_{jk}},{\bf I})$ and obtaining ${\bf {\acute F}}_{N}^{\pi_{jk}}({\bf I},\frac{1}{M}\acute{\mathbf{\Sigma}}_{\acute\pi_{jk}},{\bf I})$, given in (\ref{APPENDIX B: eq6xxx3}) and then applying Lemma 1.2, (d) and (e) follow  from applying Lemma 3 and Theorem 1, respectively, and (f) follows by defining $\delta_{jk}={\tilde f}_{N}^{\pi_{jk}}({\bf\Xi}_{jk,jk}^j,{\bf I},\frac{1}{M}{{\bf T}_j},{\bf I})$. Using the continuous mapping theorem \cite{van2000asymptotic} we obtain:
\begin{equation}
	\label{APPENDIX B: eq8}
	|{{\bf v}_{jk}}^H{\hat{\bf h}}_{jk}^j|^2-(\frac{\delta_{jk}}{1+p_{jk}\delta_{jk}})^2\stackrel{\stackrel{\text{a.s.}}{M\rightarrow\infty}}{\longrightarrow}0
\end{equation}
\subsection{Power of channel uncertainty in (\ref{Uplink_Data_Transmission_Phase_eq2})}
To calculate the average power of  ${\bf v}_{jk}^H{\tilde{\bf h}}_{jk}^j$ in (\ref{Uplink_Data_Transmission_Phase_eq2}), we note  ${\bf v}_{jk}^H{\tilde{\bf h}}_{jk}^j=(\hat{\bf h}_{jk}^j)^H{\boldsymbol\Sigma}_j{\tilde{\bf h}}_{jk}^j$ where
\begin{equation}
	\begin{split}
		\label{APPENDIX B: eq9}
		&(\hat{{\bf h}}_{jk}^j)^H{\boldsymbol{\Sigma}}_j{\tilde{{\bf h}}}_{jk}^j \stackrel{\text{(a)}}{=}\frac{({\hat{{\bf h}}}_{jk}^j)^H\mathbf{F}_{N}^{\pi_{jk}}({\bf I},\mathbf{\Sigma}_{\acute\pi_{jk}},{\bf I}) {\tilde{{\bf h}}}_{jk}^j}{1+p_{jk}({\hat{{\bf h}}}_{jk}^j)^H \mathbf{F}_{N}^{\pi_{jk}}({\bf I},\mathbf{\Sigma}_{\acute\pi_{jk}},{\bf I}){\hat{{\bf h}}}_{jk}^j}\stackrel{\text{(b)}}{\asymp}0
	\end{split}
\end{equation}
in which (a) and (b) follow from using Lemma 2 and 1.4, respectively. By the continuous mapping theorem and the dominated convergence theorem \cite{billingsley2017probability} we have
\begin{equation}
	\label{APPENDIX B: eq11}
	\mathbb{E}\{|{\bf v}_{jk}^H{\tilde{\bf h}}_{jk}^j|^2|\ {\hat{\boldsymbol{\mathcal{H}}}}_j\}\stackrel{\stackrel{\text{a.s.}}{M\rightarrow\infty}}{\longrightarrow}0
\end{equation}
\begin{figure*}
   \centering
\begin{flalign}
		\label{capGama}				
		&{\boldsymbol\Lambda}_{N,\acute N}^{\pi_{jk}}({\bf A},\mathbf{\Sigma}_{\acute\pi_{jk}},\mathbf{\Sigma}_{\acute\pi_{jk}}{\bf C}\mathbf{\Sigma}_{\acute\pi_{jk}},{\bf B})=\mathbf{\acute F}_{N}^{\pi_{jk}}({\bf A},\mathbf{\Sigma}_{\acute\pi_{jk}},{\bf C})\mathbf{\acute F}_{\acute N}^{\pi_{jk}}({\bf I},\mathbf{\Sigma}_{\acute\pi_{jk}},{\bf B})=\mathbf{\acute F}_{0}^{\pi_{jk}}({\bf A},\mathbf{\Sigma}_{\acute\pi_{jk}}{\bf C}\mathbf{\Sigma}_{\acute\pi_{jk}},\mathbf{\Sigma}_{\acute\pi_{jk}},{\bf B})\notag\\
		&-\sum_{\overset{n=1}{(ab)=\pi_{jk_n}}}^{\acute N}\frac{p_{ab}{\bf\acute F}_{n-1}^{\pi_{jk}}({\bf A},\mathbf{\Sigma}_{\acute\pi_{jk}}{\bf C}\mathbf{\Sigma}_{\acute\pi_{jk}},\mathbf{\Sigma}_{\acute\pi_{jk}},{{\bf \Upsilon}_{ab}^{jk}}){\hat{{\bf h}}}_{jk}^j({\hat{{\bf h}}}_{jk}^j)^H{\bf\acute F}_{n-1}^{\pi_{jk}}({{\bf \Upsilon}_{ab}^{jk}},\mathbf{\Sigma}_{\acute\pi_{jk}},{\bf B})}{1+p_{ab}({\hat{{\bf h}}}_{jk}^j)^H{\bf\acute F}_{n-1}^{\pi_{jk}}({{\bf \Upsilon}_{ab}^{jk}},\mathbf{\Sigma}_{\acute\pi_{jk}},{{\bf \Upsilon}_{ab}^{jk}}){\hat{{\bf h}}}_{jk}^j}\notag\\
		&-\sum_{\overset{n=1}{(ab)=\pi_{jk_n}}}^{N}\frac{p_{ab}{\bf\acute F}_{n-1}^{\pi_{jk}}({\bf A},\mathbf{\Sigma}_{\acute\pi_{jk}},{{\bf \Upsilon}_{ab}^{jk}}){\hat{{\bf h}}}_{jk}^j({\hat{{\bf h}}}_{jk}^j)^H{\bf\acute F}_{n-1}^{\pi_{jk}}({{\bf \Upsilon}_{ab}^{jk}},\mathbf{\Sigma}_{\acute\pi_{jk}},\mathbf{\Sigma}_{\acute\pi_{jk}}{\bf C}\mathbf{\Sigma}_{\acute\pi_{jk}},{\bf B})}{1+p_{ab}({\hat{{\bf h}}}_{jk}^j)^H{\bf\acute F}_{n-1}^{\pi_{jk}}({{\bf \Upsilon}_{ab}^{jk}},\mathbf{\Sigma}_{\acute\pi_{jk}},{{\bf \Upsilon}_{ab}^{jk}}){\hat{{\bf h}}}_{jk}^j}+\\
		&\sum_{\overset{n=1}{(ab)=\pi_{jk_n}}}^{N}\sum_{\overset{\acute{n}=1}{(\acute{a}\acute{b})=\pi_{jk_{\acute n}}}}^{\acute N}\frac{p_{ab}p_{\acute{a}\acute{b}}{\bf\acute F}_{n-1}^{\pi_{jk}}({\bf A},\mathbf{\Sigma}_{\acute\pi_{jk}},{{\bf \Upsilon}_{ab}^{jk}}){\hat{{\bf h}}}_{jk}^j({\hat{{\bf h}}}_{jk}^j)^H{\bf\acute F}_{n-1}^{\pi_{jk}}({{\bf \Upsilon}_{ab}^{jk}},\mathbf{\Sigma}_{\acute\pi_{jk}},{\bf C}){\bf\acute F}_{{\acute n}-1}^{\pi_{jk}}({\bf I},\mathbf{\Sigma}_{\acute\pi_{jk}},{{\bf \Upsilon}_{{\acute a}{\acute b}}^{jk}}){\hat{{\bf h}}}_{jk}^j({\hat{{\bf h}}}_{jk}^j)^H{\bf\acute F}_{{\acute n}-1}^{\pi_{jk}}({{\bf \Upsilon}_{{\acute a}{\acute b}}^{jk}},\mathbf{\Sigma}_{\acute\pi_{jk}},{\bf B})}{(1+p_{ab}({\hat{{\bf h}}}_{jk}^j)^H{\bf\acute F}_{n-1}^{\pi_{jk}}({{\bf \Upsilon}_{ab}^{jk}},\mathbf{\Sigma}_{\acute\pi_{jk}},{{\bf \Upsilon}_{ab}^{jk}}){\hat{{\bf h}}}_{jk}^j)(1+p_{{\acute a}{\acute b}}({\hat{{\bf h}}}_{jk}^j)^H{\bf\acute F}_{{\acute n}-1}^{\pi_{jk}}({{\bf \Upsilon}_{{\acute a}{\acute b}}^{jk}},\mathbf{\Sigma}_{\acute\pi_{jk}},{{\bf \Upsilon}_{{\acute a}{\acute b}}^{jk}}){\hat{{\bf h}}}_{jk}^j)}\notag
\end{flalign}
\rule{\textwidth}{0.4pt}
\begin{flalign}
		\label{capgimi}
		&{\lambda}_{N,\acute N}^{\pi_{jk}}({\bf\Xi}_{jk,jk}^j,{\bf A},\mathbf{\Sigma}_{\acute\pi_{jk}},\mathbf{\Sigma}_{\acute\pi_{jk}}{\bf C}\mathbf{\Sigma}_{\acute\pi_{jk}},{\bf B})={\tilde f}_{0}^{\pi_{jk}}({\bf\Xi}_{jk,jk}^j,{\bf A},\mathbf{\Sigma}_{\acute\pi_{jk}}{\bf C}\mathbf{\Sigma}_{\acute\pi_{jk}},\mathbf{\Sigma}_{\acute\pi_{jk}},{\bf B})\notag\\
		&-\sum_{\overset{n=1}{(ab)=\pi_{jk_n}}}^{\acute N}\frac{p_{ab}{\tilde f}_{n-1}^{\pi_{jk}}({\bf\Xi}_{jk,jk}^j,{\bf A},\mathbf{\Sigma}_{\acute\pi_{jk}}{\bf C}\mathbf{\Sigma}_{\acute\pi_{jk}},\mathbf{\Sigma}_{\acute\pi_{jk}},{{\bf \Upsilon}_{ab}^{jk}}){\tilde f}_{n-1}^{\pi_{jk}}({\bf\Xi}_{jk,jk}^j,{{\bf \Upsilon}_{ab}^{jk}},\mathbf{\Sigma}_{\acute\pi_{jk}},{\bf B})}{1+p_{ab}{\tilde f}_{n-1}^{\pi_{jk}}({\bf\Xi}_{jk,jk}^j,{{\bf \Upsilon}_{ab}^{jk}},\mathbf{\Sigma}_{\acute\pi_{jk}},{{\bf \Upsilon}_{ab}^{jk}})}\notag\\
		&-\sum_{\overset{n=1}{(ab)=\pi_{jk_n}}}^{ N}\frac{p_{ab}{\tilde f}_{n-1}^{\pi_{jk}}({\bf\Xi}_{jk,jk}^j,{\bf A},\mathbf{\Sigma}_{\acute\pi_{jk}},{{\bf \Upsilon}_{ab}^{jk}}){\tilde f}_{n-1}^{\pi_{jk}}({\bf\Xi}_{jk,jk}^j,{{\bf \Upsilon}_{ab}^{jk}},\mathbf{\Sigma}_{\acute\pi_{jk}},\mathbf{\Sigma}_{\acute\pi_{jk}}{\bf C}\mathbf{\Sigma}_{\acute\pi_{jk}},{\bf B})}{1+p_{ab}{\tilde f}_{n-1}^{\pi_{jk}}({\bf\Xi}_{jk,jk}^j,{{\bf \Upsilon}_{ab}^{jk}},\mathbf{\Sigma}_{\acute\pi_{jk}},{{\bf \Upsilon}_{ab}^{jk}})}\\
		&+\sum_{\overset{n=1}{(ab)=\pi_{jk_n}}}^{N}\sum_{\overset{\acute{n}=1}{(\acute{a}\acute{b})=\pi_{jk_{\acute n}}}}^{\acute N}\frac{p_{ab}p_{\acute{a}\acute{b}}{\tilde f}_{n-1}^{\pi_{jk}}({\bf\Xi}_{jk,jk}^j,{\bf A},\mathbf{\Sigma}_{\acute\pi_{jk}},{{\bf \Upsilon}_{ab}^{jk}}){\lambda}_{n-1,{\acute n}-1}^{\pi_{jk}}({\bf\Xi}_{jk,jk}^j,{{\bf \Upsilon}_{ab}^{jk}},\mathbf{\Sigma}_{\acute\pi_{jk}},\mathbf{\Sigma}_{\acute\pi_{jk}}{\bf C}\mathbf{\Sigma}_{\acute\pi_{jk}},{{\bf \Upsilon}_{{\acute a}{\acute b}}^{jk}}){\tilde f}_{{\acute n}-1}^{\pi_{jk}}({\bf\Xi}_{jk,jk}^j,{{\bf \Upsilon}_{{\acute a}{\acute b}}^{jk}},\mathbf{\Sigma}_{\acute\pi_{jk}},{\bf B})}{(1+p_{ab}{\tilde f}_{n-1}^{\pi_{jk}}({\bf\Xi}_{jk,jk}^j,{{\bf \Upsilon}_{ab}^{jk}},\mathbf{\Sigma}_{\acute\pi_{jk}},{{\bf \Upsilon}_{ab}^{jk}}))(1+p_{{\acute a}{\acute b}}{\tilde f}_{{\acute n}-1}^{\pi_{jk}}({\bf\Xi}_{jk,jk}^j,{{\bf \Upsilon}_{{\acute a}{\acute b}}^{jk}},\mathbf{\Sigma}_{\acute\pi_{jk}},{{\bf \Upsilon}_{{\acute a}{\acute b}}^{jk}}))}\notag
\end{flalign}\rule{\textwidth}{0.2pt}
\end{figure*}
\subsection{Power of interference in (\ref{Uplink_Data_Transmission_Phase_eq2})}
\label{APPENDIX P_int}
Consider the interference term in (\ref{Uplink_Data_Transmission_Phase_eq2}). The interference power from user $li$ is
\begin{equation}
	\label{APPENDIX B: eq12}
	\mathbb{E}\{|{\bf v}_{jk}^H{{\bf h}}_{li}^j|^2|\ {\hat{\boldsymbol{\mathcal{H}}}}_j\}=\mathbb{E}\{|({\hat{\bf h}}_{jk}^j)^H{\boldsymbol\Sigma}_j{{\bf h}}_{li}^j|^2|\ {\hat{\boldsymbol{\mathcal{H}}}}_j\}
\end{equation}
We have:
\begin{equation}
	\label{APPENDIX B: eq14}
	|({\hat{\bf h}}_{jk}^j)^H{\boldsymbol\Sigma}_j{{\bf h}}_{li}^j|^2\stackrel{\text{(a)}}{=}\frac{({\hat{\bf h}}_{jk}^j)^H{\bf \Sigma}_{jk}{\bf h}_{li}^j({\bf h}_{li}^j)^H{\boldsymbol\Sigma}_{jk}{\hat{\bf h}}_{jk}^j}{(1+p_{jk}(\hat{\bf h}_{jk}^j)^H{\boldsymbol\Sigma}_{jk}{\hat{\bf h}}_{jk}^j)^2}
\end{equation}
where (a) follows from applying Lemma 2.1. Calculating the interference power depends on the PA scheme. We consider two cases: case (1) when users $jk$ and $li$ use the same pilot sequence and hence the interference from user $li$ in (\ref{Uplink_Data_Transmission_Phase_eq2}) contributes to coherent interference; case (2) when users $jk$ and $li$ use orthogonal pilot sequences and thus the interference from user $li$ in (\ref{Uplink_Data_Transmission_Phase_eq2}) contributes to non-coherent interference. The denominator of (\ref{APPENDIX B: eq14}) is the square of the denominator of (\ref{APPENDIX B: eq6xxx1}). Therefore, our focus shifts to rewriting the numerator of (\ref{APPENDIX B: eq14}). In case (1), replacing ${{\bf h}}_{li}^j={\hat{\bf h}}_{li}^j+{\tilde{\bf h}}_{li}^j$ in (\ref{APPENDIX B: eq14}) and using the fact that ${\hat{\bf h}}_{li}^j$ and ${\tilde{\bf h}}_{li}^j$ are zero mean and uncorrelated, and then, utilizing ${\hat{\bf h}}_{li}^j={\bf \Upsilon}_{li}^{jk}{\hat{\bf h}}_{jk}^j$, following similar steps to those used in finding the desired signal power, and defining $\delta_{jlik}={\tilde f}_{N}^{\pi_{jk}}({\bf\Xi}_{jk,jk}^j,{\bf I},\frac{1}{M}{{\bf T}_j},{\bf \Upsilon}_{li}^{jk})$ we obtain
\begin{equation}
	\label{APPENDIX B: eq16}
	|{\bf v}_{jk}^H{\bf h}_{li}^j|^2-(\frac{\delta_{jlik}}{1+p_{jk}\delta_{jk}})^2\stackrel{\stackrel{\text{a.s.}}{M\rightarrow\infty}}{\longrightarrow}0
\end{equation}
In case (2), we have:
\begin{equation}
	\begin{split}
		\label{APPENDIX B: eq19}
		&(\hat{\bf h}_{jk}^j)^H{\boldsymbol\Sigma_{jk}}{\bf h}_{li}^j({\bf h}_{li}^j)^H{\boldsymbol\Sigma_{jk}}{\hat{\bf h}}_{jk}^j\\
		&\stackrel{\text{(b)}}{=}(\hat{\bf h}_{jk}^j)^H{\boldsymbol\Lambda}_{N, N}^{\pi_{jk}}({\bf I},\mathbf{\Sigma}_{\acute\pi_{jk}},\mathbf{\Sigma}_{\acute\pi_{jk}}{\bf h}_{li}^j({\bf h}_{li}^j)^H\mathbf{\Sigma}_{\acute\pi_{jk}},{\bf I}){\hat{\bf h}}_{jk}^j\\
		&\stackrel{\text{(c)}}{=}{\lambda}_{N,N}^{\pi_{jk}}({\bf\Xi}_{jk,jk}^j,{\bf I},\frac{1}{M}\acute{\mathbf{\Sigma}}_{\acute\pi_{jk}},\frac{1}{M^2}\acute{\mathbf{\Sigma}}_{\acute\pi_{jk}}{\bf h}_{li}^j({\bf h}_{li}^j)^H\acute{\mathbf{\Sigma}}_{\acute\pi_{jk}},{\bf I})
	\end{split}
\end{equation}
where (b) follows from using Lemma 2.2 to remove all channel
estimates corresponding to users who share the same pilot
as user $jk$ from ${\boldsymbol{\Sigma}_{jk}}$ and  ${\boldsymbol\Lambda}_{N, N}^{\pi_{jk}}({\bf I},\mathbf{\Sigma}_{\acute\pi_{jk}},\mathbf{\Sigma}_{\acute\pi_{jk}}{\bf h}_{li}^j({\bf h}_{li}^j)^H\mathbf{\Sigma}_{\acute\pi_{jk}},{\bf I})$ is obtained using (\ref{capGama}), and (c) follows from replacing ${\mathbf{\Sigma}_{\acute\pi_{jk}}}$ with $=\frac{1}{M}\acute{\mathbf{\Sigma}}_{\acute\pi_{jk}}$ into ${\boldsymbol\Lambda}_{N, N}^{\pi_{jk}}$ and applying Lemma 1, obtaining ${\lambda}_{N,N}^{\pi_{jk}}({\bf\Xi}_{jk,jk}^j,{\bf I},\frac{1}{M}\acute{\mathbf{\Sigma}}_{\acute\pi_{jk}},\frac{1}{M^2}\acute{\mathbf{\Sigma}}_{\acute\pi_{jk}}{\bf h}_{li}^j({\bf h}_{li}^j)^H\acute{\mathbf{\Sigma}}_{\acute\pi_{jk}},{\bf I})$ given in (\ref{capgimi}). The remainder of the proof involves examining each term within ${\lambda}_{N,N}^{\pi_{jk}}$ in (\ref{APPENDIX B: eq19}), and systematically applying the recursive equations from Lemma 4 to ${\tilde f}_{2}^{\pi_{jk}}$, ${\tilde f}_{3}^{\pi_{jk}}$,…,${\tilde f}_{N}^{\pi_{jk}}$, iteratively unfolding them until we ultimately reach ${\tilde f}_{1}^{\pi_{jk}}$. At this stage, 
${\lambda}_{N,N}^{\pi_{jk}}$ in (\ref{APPENDIX B: eq19}) is expressed solely in terms of ${\tilde f}_{0}^{\pi_{jk}}$ and ${\tilde f}_{1}^{\pi_{jk}}$. It is important to note that some of these ${\tilde f}_{0}^{\pi_{jk}}$  and ${\tilde f}_{1}^{\pi_{jk}}$ functions depend on the random product  ${\bf h}_{li}^j({\bf h}_{li}^j)^H$, while the others do not. The ${\tilde f}_{1}^{\pi_{jk}}$ functions that are independent of ${\bf h}_{li}^j({\bf h}_{li}^j)^H$ can be expressed in terms of ${\tilde f}_{0}^{\pi_{jk}}$ using Lemma 4. We then turn our attention to the ${\tilde f}_{0}^{\pi_{jk}}$ and ${\tilde f}_{1}^{\pi_{jk}}$  functions that do depend on ${\bf h}_{li}^j({\bf h}_{li}^j)^H$. Generally, the ${\tilde f}_{0}^{\pi_{jk}}$ functions that depend on ${\bf h}_{li}^j({\bf h}_{li}^j)^H$ can be reformulated as follows:
\begin{equation}
	\begin{split}
		\label{APPENDIX B: eejq19}
		&{\tilde f}_{0}^{\pi_{jk}}({\bf\Xi}_{jk,jk}^j,{{\bf \Upsilon}_{ab}^{jk}},\frac{1}{M}\acute{\mathbf{\Sigma}}_{\acute\pi_{jk}},\frac{1}{M^2}\acute{\mathbf{\Sigma}}_{\acute\pi_{jk}}{\bf h}_{li}^j({\bf h}_{li}^j)^H\acute{\mathbf{\Sigma}}_{\acute\pi_{jk}},{{\bf \Upsilon}_{{\acute a}{\acute b}}^{jk}})=\\
		&({\bf h}_{li}^j)^H\mathbf{\Sigma}_{\acute\pi_{jk}}{\bf\Xi}_{{\acute a}{\acute b},ab}^j\mathbf{\Sigma}_{\acute\pi_{jk}}{\bf h}_{li}^j \stackrel{\text{(d)}}{=}\\
		&(\hat{\bf h}_{li}^j)^H{\boldsymbol\Lambda}_{N, N}^{\acute\pi_{jk}\cup\acute\pi_{li}}({\bf I},\frac{1}{M}\acute{\mathbf{\Sigma}}_{\acute\pi_{jk}\cup\acute\pi_{li}},\frac{1}{M^2}\acute{\mathbf{\Sigma}}_{\acute\pi_{jk}\cup\acute\pi_{li}}{\bf\Xi}_{{\acute a}{\acute b},ab}^j\acute{\mathbf{\Sigma}}_{\acute\pi_{jk}\cup\acute\pi_{li}},{\bf I}){\hat{\bf h}}_{li}^j\\
		&\stackrel{\text{(e)}}{=}{\lambda}_{N,N}^{\acute\pi_{jk}\cup\acute\pi_{li}}({\bf\Xi}_{li,li}^j,{\bf I},\frac{1}{M}\acute{\mathbf{\Sigma}}_{\acute\pi_{jk}\cup\acute\pi_{li}},\frac{1}{M^2}\acute{\mathbf{\Sigma}}_{\acute\pi_{jk}\cup\acute\pi_{li}}{\bf\Xi}_{{\acute a}{\acute b},ab}^j\acute{\mathbf{\Sigma}}_{\acute\pi_{jk}\cup\acute\pi_{li}},{\bf I})
	\end{split}
\end{equation}
where (d) follows from replacing ${{\bf h}}_{li}^j={\hat{\bf h}}_{li}^j+{\tilde{\bf h}}_{li}^j$ and applying Lemma 2 to remove $\hat{\bf h}_{li}^j$ as well as all channel estimates corresponding to  users who share the same pilot as user $li$ from ${\boldsymbol{\Sigma}_{jk}}$, and replacing ${\mathbf{\Sigma}_{\acute\pi_{jk}\cup\acute\pi_{li}}}$ with $\frac{1}{M}\acute{\mathbf{\Sigma}}_{\acute\pi_{jk}\cup\acute\pi_{li}}$, and (e) follows from applying Lemma \ref{Lemma: Appendix A lem1}. The ${\tilde f}_{1}^{\pi_{jk}}$ functions that depend on ${\bf h}_{li}^j({\bf h}_{li}^j)^H$ can generally be reformulated using (\ref{R_1_equations}) and (\ref{R_2_equations}), where the right-hand sides of the equations no longer depend on ${\bf h}_{li}^j({\bf h}_{li}^j)^H$. Once the rewriting of all terms in ${\lambda}_{N,N}^{\pi_{jk}}$ from (\ref{APPENDIX B: eq19}) is complete, we proceed by examining each term within ${\lambda}_{N,N}^{\acute\pi_{jk}\cup\acute\pi_{li}}$, as introduced in (\ref{APPENDIX B: eejq19}), (\ref{R_1_equations}), and (\ref{R_2_equations}), and applying Lemma 4 iteratively until we reach the ${\tilde f}_{0}^{\pi_{jk}\cup\pi_{li}}$ functions, which are dependent on ${\bf\Xi}_{{\acute a}{\acute b},ab}^j$, ${\bf\Xi}_{{\acute a}{\acute b},mr}^j$, and ${\bf\Xi}_{mr,ab}^j$ in general form. Subsequently, we apply Theorem 1 and 2, specifying  $\mathbf{\Theta}={\bf\Xi}_{{\acute a}{\acute b},ab}^j$, $\mathbf{\Theta}={\bf\Xi}_{{\acute a}{\acute b},mr}^j$, and $\mathbf{\Theta}={\bf\Xi}_{mr,ab}^j$ in the general form, and $\mathbf{\Theta}={\bf\Xi}_{jk,li}^j$ with $j k,li\in\{jk,ab,mr\}$ for the inputs of (\ref{APPENDIX B: eejq19}), (\ref{R_1_equations}), and (\ref{R_2_equations}) for cases other than the general form, to obtain the final results:
\begin{equation}
	\begin{split}
		\label{APPENDIX B: case2}
		&{\lambda}_{N,N}^{\pi_{jk}}({\bf\Xi}_{jk,jk}^j,{\bf I},\frac{1}{M}\acute{\mathbf{\Sigma}}_{\acute\pi_{jk}},\frac{1}{M^2}\acute{\mathbf{\Sigma}}_{\acute\pi_{jk}}{\bf h}_{li}^j({\bf h}_{li}^j)^H\acute{\mathbf{\Sigma}}_{\acute\pi_{jk}},{\bf I})\stackrel{\text{(f)}}{\asymp}\mu_{jlik}
	\end{split}
\end{equation}
where (f) follows from defining $\mu_{jlik}={\lambda}_{N,N}^{\pi_{jk}}({\bf\Xi}_{jk,jk}^j,{\bf I},\frac{1}{M}{\bf T}_j,\frac{1}{M^2}\acute{\bf T}_j,{\bf I})$.
\begin{figure*}
   \centering
\begin{equation}
	\begin{split}
		\label{R_1_equations}
		&{{\tilde f}}_{1}^{\pi_{jk}}({\bf\Xi}_{jk,jk}^j,{{\bf \Upsilon}_{ab}^{jk}},\frac{1}{M}\acute{\mathbf{\Sigma}}_{\acute\pi_{jk}\cup\acute\pi_{li}},\frac{1}{M^2}\acute{\mathbf{\Sigma}}_{\acute\pi_{jk}\cup\acute\pi_{li}}{\bf h}_{li}^j({\bf h}_{li}^j)^H\acute{\mathbf{\Sigma}}_{\acute\pi_{jk}\cup\acute\pi_{li}}, {{\bf \Upsilon}_{{\acute a}{\acute b}}^{jk}})={\lambda}_{N,N}^{\acute\pi_{jk}\cup\acute\pi_{li}}({\bf\Xi}_{li,li}^j,{\bf I},\frac{1}{M}\acute{\mathbf{\Sigma}}_{\acute\pi_{jk}\cup\acute\pi_{li}},\frac{1}{M^2}\acute{\mathbf{\Sigma}}_{\acute\pi_{jk}\cup\acute\pi_{li}}{\bf\Xi}_{{\acute a}{\acute b},ab}^j\acute{\mathbf{\Sigma}}_{\acute\pi_{jk}\cup\acute\pi_{li}}, {\bf I})\\
		&-\frac{p_{mr}{{\tilde f}}_{0}^{\pi_{jk}}({\bf\Xi}_{jk,jk}^j,{{\bf \Upsilon}_{ab}^{jk}},\frac{1}{M}\acute{\mathbf{\Sigma}}_{\acute\pi_{jk}\cup\acute\pi_{li}},{\bf \Upsilon}_{mr}^{jk}){\lambda}_{N,N}^{\acute\pi_{jk}\cup\acute\pi_{li}}({\bf\Xi}_{li,li}^j,{\bf I},\frac{1}{M}\acute{\mathbf{\Sigma}}_{\acute\pi_{jk}\cup\acute\pi_{li}},\frac{1}{M^2}\acute{\mathbf{\Sigma}}_{\acute\pi_{jk}\cup\acute\pi_{li}}{\bf\Xi}_{{\acute a}{\acute b},mr}^j\acute{\mathbf{\Sigma}}_{\acute\pi_{jk}\cup\acute\pi_{li}}, {\bf I})}{1+{\hat{p}}_{mr}{ {\tilde f}}_{0}^{\pi_{jk}}({\bf\Xi}_{jk,jk}^j,{\bf \Upsilon}_{mr}^{jk},\frac{1}{M}\acute{\mathbf{\Sigma}}_{\acute\pi_{jk}\cup\acute\pi_{li}}, {\bf \Upsilon}_{mr}^{jk})}
	\end{split}
\end{equation}
\rule{\textwidth}{0.2pt}
\begin{equation}
	\begin{split}
		\label{R_2_equations}
		&{{\tilde f}}_{1}^{\pi_{jk}}({\bf\Xi}_{jk,jk}^j,{{\bf \Upsilon}_{ab}^{jk}},\frac{1}{M^2}\acute{\mathbf{\Sigma}}_{\acute\pi_{jk}\cup\acute\pi_{li}}{\bf h}_{li}^j({\bf h}_{li}^j)^H\acute{\mathbf{\Sigma}}_{\acute\pi_{jk}\cup\acute\pi_{li}},\frac{1}{M}\acute{\mathbf{\Sigma}}_{\acute\pi_{jk}\cup\acute\pi_{li}},{{\bf \Upsilon}_{{\acute a}{\acute b}}^{jk}})={\lambda}_{N,N}^{\acute\pi_{jk}\cup\acute\pi_{li}}({\bf\Xi}_{li,li}^j,{\bf I},\frac{1}{M}\acute{\mathbf{\Sigma}}_{\acute\pi_{jk}\cup\acute\pi_{li}},\frac{1}{M^2}\acute{\mathbf{\Sigma}}_{\acute\pi_{jk}\cup\acute\pi_{li}}{\bf\Xi}_{{\acute a}{\acute b},ab}^j\acute{\mathbf{\Sigma}}_{\acute\pi_{jk}\cup\acute\pi_{li}}, {\bf I})\\
		&-\frac{p_{mr}{\lambda}_{N,N}^{\acute\pi_{jk}\cup\acute\pi_{li}}({\bf\Xi}_{li,li}^j,{\bf I},\frac{1}{M}\acute{\mathbf{\Sigma}}_{\acute\pi_{jk}\cup\acute\pi_{li}},\frac{1}{M^2}\acute{\mathbf{\Sigma}}_{\acute\pi_{jk}\cup\acute\pi_{li}}{\bf\Xi}_{mr,ab}^j\acute{\mathbf{\Sigma}}_{\acute\pi_{jk}\cup\acute\pi_{li}}, {\bf I}){{\tilde f}}_{0}^{\pi_{jk}}({\bf\Xi}_{jk,jk}^j,{{\bf \Upsilon}_{mr}^{jk}},\frac{1}{M}\acute{\mathbf{\Sigma}}_{\acute\pi_{jk}\cup\acute\pi_{li}},{\bf \Upsilon}_{{\acute a}{\acute b}}^{jk})}{1+{\hat{p}}_{mr}{ {\tilde f}}_{0}^{\pi_{jk}}({\bf\Xi}_{jk,jk}^j,{\bf \Upsilon}_{mr}^{jk},\frac{1}{M}\acute{\mathbf{\Sigma}}_{\acute\pi_{jk}\cup\acute\pi_{li}}, {\bf \Upsilon}_{mr}^{jk})}
	\end{split}
\end{equation}\rule{\textwidth}{0.2pt}\end{figure*}
\subsection{power of noise in (\ref{Uplink_Data_Transmission_Phase_eq2})}
To calculate the power of noise in (\ref{Uplink_Data_Transmission_Phase_eq2}), we have 
\begin{equation}
	\begin{split}
		\label{APPENDIX B: eq331}
		\|\textit{\bf v}_{jk} \|^2=|({\hat{\bf h}}_{jk}^j)^H{\boldsymbol\Sigma}_j|^2\stackrel{\text{(a)}}{=}\frac{({\hat{\bf h}}_{jk}^j)^H{\boldsymbol\Sigma}_{jk}{\boldsymbol\Sigma}_{jk}{\hat{\bf h}}_{jk}^j}{(1+p_{jk}({\hat{\bf h}}_{jk}^j)^H{\boldsymbol\Sigma}_{jk}{\hat{\bf h}}_{jk}^j)^2}
	\end{split}
\end{equation}
where (a) follows from applying Lemma 2.1. The denominator of (\ref{APPENDIX B: eq331}) is the square of the denominator of (\ref{APPENDIX B: eq6xxx1}). Therefore, our focus shifts to rewriting the numerator of (\ref{APPENDIX B: eq331}). We have:
\begin{equation}
	\begin{split}
		\label{APPENDIX B: eq31}
		&({\hat{\bf h}}_{jk}^j)^H{\boldsymbol\Sigma}_{jk}{\boldsymbol\Sigma}_{jk}{\hat{\bf h}}_{jk}^j\stackrel{\text{(b)}}{=}\\
		&(\hat{\bf h}_{jk}^j)^H{\boldsymbol\Lambda}_{N, N}^{\pi_{jk}}({\bf I},\mathbf{\Sigma}_{\acute\pi_{jk}},\mathbf{\Sigma}_{\acute\pi_{jk}}\mathbf{\Sigma}_{\acute\pi_{jk}},{\bf I}){\hat{\bf h}}_{jk}^j\stackrel{\text{(c)}}{=}\\
		&{\lambda}_{N,N}^{\pi_{jk}}({\bf\Xi}_{jk,jk}^j,{\bf I},\frac{1}{M}\acute{\mathbf{\Sigma}}_{\acute\pi_{jk}},\frac{1}{M^2}\acute{\mathbf{\Sigma}}_{\acute\pi_{jk}}\acute{\mathbf{\Sigma}}_{\acute\pi_{jk}},{\bf I})\stackrel{\text{(d)}}{\asymp}\delta_{jk}^{\prime\prime}
	\end{split}
\end{equation}
where (b) follows from applying Lemma 2, (c) follows from replacing ${\mathbf{\Sigma}_{\acute\pi_{jk}}}$ with $\frac{1}{M}\acute{\mathbf{\Sigma}}_{\acute\pi_{jk}}$, and applying Lemma 1, and (d) follows applying Lemma 3, and Theorems 1 and 2, and defining $\delta_{jk}^{\prime\prime}={\lambda}_{N,N}^{\pi_{jk}}({\bf\Xi}_{jk,jk}^j,{\bf I},\frac{1}{M}{\bf T}_j,\frac{1}{M^2}\mathbf{T}_j^{\prime\prime},{\bf I})$.

\vfill
\end{document}